\documentclass[12pt]{article}

\usepackage{latexsym}
\usepackage{amssymb,amsfonts,amsmath}
\usepackage{graphicx} 
\usepackage{indentfirst}
\usepackage{bbm}
\usepackage{amssymb}
\usepackage{verbatim}
\usepackage{amsmath, amsthm,amssymb}
\usepackage{mathrsfs}
\usepackage{hyperref}
\usepackage{amsfonts}
\usepackage{dsfont}
\usepackage{cite}
\usepackage{xcolor}
\usepackage[multiple]{footmisc}
\usepackage{ytableau}
\usepackage{tikz}

\parskip=\medskipamount

\newcommand {\cD}{{\cal D}}
\newcommand {\cE}{{\cal E}}

\newcommand {\cK}{{\cal K}}

\newcommand {\cM}{{\cal M}}
\newcommand {\cN}{{\cal N}}

\newcommand {\cR}{{\cal R}}
\newcommand {\cS}{{\cal S}}
\newcommand {\cT}{{\cal T}}
\newcommand {\cU}{{\cal U}}

\def\a{\alpha}

\def\c{\chi}
\def\d{\delta}

\def\f{\phi}
\def\g{\gamma}
\def\G{\Gamma}

\def\j{\psi}

\def\l{\lambda}
\def\m{\mu}

\def\o{\omega}
\def\p{\pi}
\def\q{\theta}
\def\r{\rho}
\def\s{\sigma}

\def\x{\xi}
\def\z{\zeta}
\def\D{\Delta}
\def\F{\Phi}
\def\J{\Psi}
\def\L{\Lambda}
\def\O{\Omega}

\def\S{\Sigma}

\def\X{\Xi}

\def\rd{{\rm d}}
\def\ri{{\rm i}}
\def\re{{\rm e}}

\newcommand{\ve}{\varepsilon}                            

\newcommand{\pa}{\partial}                           
\newcommand{\hf}{\frac12}

\newcommand{\vf}{\varphi}

\newcommand{\be}{\begin{equation}}
\newcommand{\ee}{\end{equation}}
\newcommand{\bea}{\begin{eqnarray}}
\newcommand{\eea}{\end{eqnarray}}
\newcommand{\non}{\nonumber}
\newcommand{\bm}[1]{\mbox{\boldmath$#1$}}

\def\double #1{#1{\hbox{\kern-2pt $#1$}}}

\newcommand{\OI}{{\overline{I}}}
\newcommand{\OJ}{{\overline{J}}}
\newcommand{\OK}{{\overline{K}}}
\newcommand{\OL}{{\overline{L}}}
\newcommand{\OM}{{\overline{M}}}

\newcommand{\UI}{{\underline{I}}}
\newcommand{\UJ}{{\underline{J}}}
\newcommand{\UK}{{\underline{K}}}
\newcommand{\UL}{{\underline{L}}}
\newcommand{\UM}{{\underline{M}}}

\newif\ifdtup

\newcommand{\bsubeq}{\begin{subequations}}
\newcommand{\esubeq}{\end{subequations}}

\newcommand{\1}{{\underline{1}}}

\newcommand{\Iu}{\underline{I}}

\newcommand{\Ju}{\underline{J}}

\newcommand{\Io}{{\overline{I}}}

\newcommand{\Jo}{\overline{J}}

\numberwithin{equation}{section}

\newcommand{\sSU}{\mathsf{SU}}
\newcommand{\sSL}{\mathsf{SL}}

\newcommand{\sSO}{\mathsf{SO}}
\newcommand{\sU}{\mathsf{U}}

\newcommand{\sOSp}{\mathsf{OSp}}

\begin{document}

\begin{titlepage}
\begin{flushright}
November, 2022 \\
Revised version: December, 2022
\end{flushright}
\vspace{5mm}

\begin{center}
{\Large \bf 
Conformal $(p,q)$ supergeometries in two dimensions}
\end{center}

\begin{center}	
{\bf Sergei M. Kuzenko and Emmanouil S. N. Raptakis} \\
\vspace{5mm}

\footnotesize{
{\it  Department of Physics M013, The University of Western Australia\\
35 Stirling Highway, Perth W.A. 6009, Australia}}  
~\\
\vspace{2mm}
~\\
Email: \texttt{ 
	sergei.kuzenko@uwa.edu.au, emmanouil.raptakis@research.uwa.edu.au
}\\
\vspace{2mm}

\end{center}

\begin{abstract}
\baselineskip=14pt
We propose a superspace formulation for conformal $(p,q)$ supergravity 
in two dimensions as a gauge theory of the superconformal group 
$\mathsf{OSp}_0 (p|2; {\mathbb R} ) \times  \mathsf{OSp}_0 (q|2; {\mathbb R} )$ with a flat connection.
Upon degauging of certain local symmetries, this conformal superspace is shown to reduce to a conformally flat $\mathsf{SO}(p) \times \mathsf{SO}(q)$ superspace with the following properties: (i) its structure group is a direct product of the Lorentz group 
and $\mathsf{SO}(p) \times \mathsf{SO}(q)$; and (ii) the residual local scale symmetry is realised by super-Weyl transformations with an unconstrained real parameter.  As an application of the formalism, we describe ${\cal N}$-extended AdS superspace as a maximally symmetric supergeometry in the $p=q \equiv \cal N$ case. 
If at least one of the parameters $p$ or $q$ is even, alternative superconformal groups and, thus, conformal superspaces exist. 
In particular, if $p = 2n$, a possible choice of 
the superconformal group is  $\mathsf{SU}(1,1|n) \times \mathsf{OSp}_0 (q|2; {\mathbb R} )$, for $n \neq 2$, and $\mathsf{PSU}(1,1|2) \times \mathsf{OSp}_0 (q|2; {\mathbb R} )$, when $n=2$. 
In general, a conformal superspace formulation is associated with a supergroup 
$  G = G_L \times G_R$, where the simple supergroups  $G_L$ and $G_R$ can be any of the extended superconformal groups, which were 
classified by G\"unaydin, Sierra and Townsend.
Degauging the corresponding conformal superspace leads to a conformally flat $H_L \times H_R$ superspace, where $H_L $ ($H_R$) is the $R$-symmetry
subgroup of $G_L$ ($G_R$). Additionally, for  the $p,q \leq 2$ cases we propose composite primary multiplets which generate the Gauss-Bonnet invariant and supersymmetric extensions of the Fradkin-Tseytlin term.
\end{abstract}
\vspace{5mm}

\vfill

\vfill
\end{titlepage}

\newpage
\renewcommand{\thefootnote}{\arabic{footnote}}
\setcounter{footnote}{0}

\tableofcontents{}
\vspace{1cm}
\bigskip\hrule

\allowdisplaybreaks


\section{Introduction}

Local superconformal symmetry has played a major role in string theory and supergravity. The $\cN=1$ spinning string can be formulated as a two-dimensional (2D) linear sigma model coupled either to $\cN=1$ Poincar\'e supergravity 
\cite{DeserZumino,BDVH} with super-Weyl invariance \cite{Howe1979}
or to $\cN=1$ conformal supergravity \cite{vanNieuwenhuizen:1985an}. 
Similarly, the $\cN=2$ spinning string can be realised as a Weyl invariant matter-coupled $\cN=2$ supergravity theory \cite{BrinkSchwarz} or as  $\cN=2$ conformal supergravity coupled to a linear sigma model  \cite{vanNieuwenhuizen:1985an}. 

In the component setting, 
conformal $(p,q)$ supergravity in two dimensions was described as a gauge theory of the superconformal algebra
$\mathfrak{osp} (p|2; {\mathbb R} ) \times  \mathfrak{osp} (q|2; {\mathbb R} )$,
for $p,q \leq 2$, in the mid 1980s \cite{vanNieuwenhuizen:1985an, Uematsu:1984zy, Uematsu:1986de, Hayashi:1986ev, Uematsu:1986aa, Bergshoeff:1985qr, Bergshoeff:1985gc, McCabe:1986jg}.
Pernici and van Nieuwenhuizen \cite{Pernici:1985dq} constructed $\cN=4 \equiv (4,4)$ conformal supergravity as a gauge theory of the superconformal algebra
$\mathfrak{psu} (1,1|2 ) \times  \mathfrak{psu} (1,1|2 )$. They coupled $\cN=4$ conformal supergravity to an arbitrary number of $\cN=4$ scalar multiplets.
An alternative approach was put forward by Schoutens \cite{Schoutens:1986kz} who formulated $d=2$ conformal supergravity with 
$\cN=0, 1,2$ and 4 as gauge theories corresponding to infinite-dimensional superalgebras.

In this paper, as a generalisation of the component results, we propose superspace formulations for conformal $(p,q)$ supergravity theories 
in two dimensions for arbitrary $p $ and $q$.
We mostly concentrate on constructing conformal $(p,q)$ supergravity 
as a gauge theory of the superconformal group 
$\mathsf{OSp}_0 (p|2; {\mathbb R} ) \times  \mathsf{OSp}_0 (q|2; {\mathbb R} )$ with a flat connection.\footnote{We point out that $\mathsf{OSp}_0 (p|2; {\mathbb R} ) \times  \mathsf{OSp}_0 (q|2; {\mathbb R} )$ is the connected superconformal group of 
compactified $(p,q)$ Minkowski superspace in two dimensions, eq. \eqref{B.1}, see \cite{KT-M2021} for  the technical details. Strictly speaking, if $p$ is even, 
$\mathsf{OSp}_0 (p|2; {\mathbb R} ) $ should be replaced with 
$\mathsf{OSp}_0 (p|2; {\mathbb R} ) /{\mathbb Z}_2$, and similar for $q$. 
However, we will not pay attention to such technical details.
}
However, our approach allows one to construct a conformal superspace formulation that is  associated with a supergroup 
$  G = G_L \times G_R$, where the simple supergroups  $G_L$ and $G_R$ can be any of the extended superconformal groups, which were 
classified by G\"unaydin, Sierra and Townsend \cite{GST} (see also \cite{McCabe:1986vb}).
Our $d=2$ construction  is a natural extension of the conformal superspace approaches in higher dimensions $3 \leq d \leq 6$ 
\cite{ButterN=1, ButterN=2, BKNT-M1, BKNT-M3, BKNT}.
From the conceptual point of view, these approaches are superspace analogues 
of the formulation for conformal gravity as the gauge theory of the conformal group
in four dimensions  \cite{KTvN1}.

It is appropriate to give a few comments about conformal gravity in $d$ dimensions following the 
discussions in \cite{FVP,BKNT-M1,BKNT}.
Beyond three dimensions, $d>3$,  the algebra of conformal covariant derivatives is
\begin{align}
[\nabla_a,\nabla_b]=-\frac{1}{2}C_{abcd}M^{cd}-\frac{1}{2(d-3)}\nabla^dC_{abcd}K^c~,
\qquad d>3~,
\end{align}
with $M^{cd}$ and $K^c$ being the Lorentz and special conformal generators, respectively.
It is determined by the Weyl tensor $C_{abcd}$, 
see appendix \ref{sectionA.1} for its properties.
For $d=3$ the algebra of conformal covariant derivatives looks simpler
\begin{align}
[\nabla_a,\nabla_b]=-\frac{1}{2}\ve_{abc}W^{cd}K_d~, \qquad d=3~,
\end{align}
where $W_{ab}$ is the Cotton tensor, see appendix \ref{sectionA.1} for its properties. Finally, in the $d=2$ case the conformal connection is flat, 
\begin{align}
[\nabla_a,\nabla_b]=0~, \qquad d=2~.
\label{1.4}
\end{align}

Actually,  the $d=2$ case is somewhat subtle. If one gauges the $d=2$ conformal group
$\mathsf{SL} (2 ,{\mathbb R} ) \times \mathsf{SL} (2, {\mathbb R} ) $
and imposes the same constraints as in higher dimensions \cite{KTvN1} (see also \cite{FVP,BKNT-M1,BKNT} for a review), then the  resulting algebra of conformal covariant derivatives turns out to be 
\begin{align}
[\nabla_a,\nabla_b]=\ve_{ab} W^c K_c~, 
\label{1.5}
\end{align}
where the special conformal curvature $W^c$ is a primary field, as discussed in section 
\ref{section3}. However, with $W^c \neq 0$ the theory is not equivalent to conformal gravity, since there is an additional gauge field along with the gravitational field. That is why one is forced to impose the conformal flatness condition \eqref{1.4}.
As a result, the special conformal connection $\mathfrak{f}_{ab}$ (which corresponds to the special conformal generator) 
becomes a non-local function of the vielbein (in a gauge where the dilatation connection $b_a$ is gauged away). This is in contrast to the situation in $d>2$ spacetime dimensions, reviewed in appendix \ref{AppendixA}, where $\mathfrak{f}_{ab}$ is proportional to the Schouten tensor, eq. \eqref{SCconn}.

The lessons from $d=2$ conformal gravity lead us to define
$d=2$ conformal $(p,q)$ supergravity 
as a gauge theory of the superconformal group 
$\mathsf{OSp}_0 (p|2; {\mathbb R} ) \times  \mathsf{OSp}_0 (q|2; {\mathbb R} )$ with a flat connection.
Upon degauging of certain local symmetries, this conformal superspace is shown to reduce to a conformally flat  superspace with its structure group being a direct product of the Lorentz group and $\mathsf{SO}(p) \times \mathsf{SO}(q)$.
The conformal flatness of the resulting  $\mathsf{SO}(p) \times \mathsf{SO}(q)$ superspace means that its supergeometry is  describable locally by a single prepotential modulo purely gauge degrees of freedom. 

An important comment is in order. In the $d=2$ case one can try to follow the philosophy of \cite{ButterN=1, ButterN=2, BKNT-M1, BKNT-M3, BKNT} to construct conformal superspace formulations in higher dimensions $3\leq d \leq 6$, which is:
the  curvature structure of conformal superspace should resemble that of the super Yang-Mills one. Then one will end up with a conformal $(p,q)$ superspace with non-vanishing curvature as an extension of the non-supersymmetric geometry \eqref{1.5}.
This idea will be elaborated in appendix \ref{AppendixC} for 
$(1,0)$ supersymmetry.

This paper is organised as follows. Section \ref{Section2} is devoted to a derivation of the infinite-dimensional superconformal algebra of Minkowski superspace 
$\mathbb{M}^{(2|p,q)}$ by employing its conformal Killing supervector fields. 
We then describe the additional constraints of the conformal Killing supervector fields 
which single out the finite-dimensional superconformal algebra $\mathfrak{osp}(p|2;\mathbb{R}) \oplus \mathfrak{osp}(q|2;\mathbb{R})$.
In section \ref{section3} we review the formulation of conformal gravity in two dimensions as the gauge theory of 
the $d=2$ conformal group
$\mathsf{SL} (2 ,{\mathbb R} ) \times \mathsf{SL} (2, {\mathbb R} ) $.
Building on the construction of conformal gravity,  in section \ref{Section4} 
we formulate conformal $(p,q)$ supergravity 
in two dimensions as a gauge theory of the superconformal group 
$\mathsf{OSp}_0 (p|2; {\mathbb R} ) \times  \mathsf{OSp}_0 (q|2; {\mathbb R} )$ with a flat connection.
The procedure of `degauging' from this superconformal formulation to the $\sSO(p) \times \sSO(q)$ superspaces is described in section \ref{Section5}. 
Section \ref{Section6} is mostly devoted to generalisations of the results derived in
section \ref{Section5}. 
Such generalisations become possible in the case that at least one of the parameters $p$ or $q$ is even, and then alternative superconformal groups and, thus, conformal superspaces exist. 
The main body of this paper is accompanied by three technical appendices. 
Appendix \ref{AppendixA} reviews conformal geometry in $d \geq 3$ dimensions following \cite{BKNT-M1,BKNT}.
In appendix \ref{AppendixC} we construct $\mathcal{N}=(1,0)$ conformal superspace with non-vanishing curvature.
Appendix \ref{AppendixB} is devoted to the supertwistor realisation of 
compactified Minkowski superspace $\overline{\mathbb M}^{(2|2n,q)}$
as a homogeneous space of the superconformal group
$ \mathsf{SU} (1,1|n )
\times  {\sOSp}_0 (q|2; {\mathbb R} )$.

Throughout this paper we will make use of two types of notation, $(p,q)$ and $\cN=(p,q)$, to denote superspaces with $p$ left and $q$ right real spinor coordinates. Additionally, the special case $p=q$ is also referred to as  $\cN = p$.


\section{The conformal Killing supervector fields of 
$\mathbb{M}^{(2|p,q)}$}
\label{Section2}

In the case of a $d$-dimensional superconformal field theory in Minkowski superspace ${\mathbb M}^{d|\d}$, 
its symmetries are formulated in terms of the conformal Killing supervector fields of  ${\mathbb M}^{d|\d}$.
 This section is devoted the description of the conformal Killing supervector fields of 
$\mathbb{M}^{(2|p,q)}$, 
the $(p,q)$ Minkowski superspace in two dimensions \cite{Hull:1985jv}.

Minkowski superspace $\mathbb{M}^{(2|p,q)}$
is  parametrised by the real coordinates $z^{A} = (x^{a},\q^{+ \Io},\q^{- \Iu})$, where $x^{a} = (x^{++},x^{--})= \frac{1}{\sqrt 2}(x^0 + x^1, x^0 - x^1)$, $\Io = \overline{1}, \dots , \overline{p}$ and $\Iu = \1, \dots , \underline{q}$. 
Its covariant derivatives $D_A = (\partial_{a}, D_+^{\Io}, D_-^{\Iu})$ take the form
\bea
\pa_{a}:=\frac{\pa}{\pa x^{a}}= ( \pa_{++} , \pa_{--} )~, \quad
D_{+}^{\Io}
:=
\frac{\pa}{\pa\q^{+ \Io}}
+ \ri \q^{+ \Io} \pa_{++}
~,\quad
D_{-}^{\Iu}
:=
\frac{\pa}{\pa\q^{- \Iu}}
+ \ri \q^{- \Iu}\pa_{--}
~,
\eea
and satisfy the algebra:
\bea
\{ D_{+}^{\Io} , D_{+}^{\Jo} \} =  2 \ri \d^{\Io \Jo} \partial_{++} ~, \qquad  
\{ D_{-}^{\Iu} , D_{-}^{\Ju} \} =  2 \ri \d^{\Iu \Ju} \partial_{--} ~.
\eea
We emphasise that for $p=0$ the left spinor covariant derivative $D_+^{\OI}$ does not appear, similarly $D_-^\UI$ is not present for $q=0$.

The conformal Killing supervector fields of $\mathbb{M}^{(2|p,q)}$,
\be 
\label{2.3}
\xi = \xi^{a} \partial_{a} + \xi^{+ \Io} D_+^{\Io} + \xi^{- \Iu} D_-^{\Iu} = \xi^{++} \partial_{++} + \xi^{--} \partial_{--} + \xi^{+ \Io} D_+^{\Io} + \xi^{- \Iu} D_-^{\Iu} = \bar{\xi} ~,
\ee
may be defined to satisfy the constraints
\be 
\label{2.4}
[\xi , D_+^{\Io} ] = - (D_+^{\Io} \xi^{+ \Jo}) D_+^{\Jo} ~, \qquad
[\xi , D_-^{\Iu} ] = - (D_-^{\Iu} \xi^{- \Ju}) D_-^{\Ju} ~.
\ee
We note that for vanishing $p$ ($q$), the spinor $\xi^{+ \OI}$ ($\xi^{- \UI}$) must be turned off.
From $\eqref{2.4}$ we obtain the fundamental equations
\begin{subequations}\label{2.5}
\bea
D_+^{\Io} \xi^{--} &=& 0 \quad \implies \quad \pa_{++} \x^{--}=0~,
 \label{2.5a}\\
D_-^{\Iu} \xi^{++} &=& 0  \quad \implies \quad \pa_{--} \x^{++}=0~, \label{2.5b}
\eea
\end{subequations}
and expressions for the spinor parameters
\begin{align}
\label{2.6}
\xi^{+ \Io} = - \frac{\ri}{2} D_+^{\Io} \xi^{++} ~, \qquad \xi^{- \Iu} = - \frac{\ri}{2} D_-^{\Iu} \xi^{--}~.
\end{align}
Hence, we see that every conformal Killing supervector field \eqref{2.3} is completely determined by its vector parameter $\xi^{a}$. Additionally, the equations \eqref{2.5} tell us  that
\begin{align}
	\xi^{++} =\xi^{++}(x^{++}, \q^+ )	
	 \equiv  \xi^{++}(\z_L)  ~, \qquad 
	 \xi^{--} =  \xi^{--}(x^{--}, \q^- )   \equiv \xi^{--}(\z_R) ~.
\end{align}
Here $\xi^{++}(\z_L)$ and $\xi^{--}(\z_R)$ are arbitrary functions of $\z_L$ and $\z_R$, respectively.
 
Taking  \eqref{2.5} into account, the equations \eqref{2.4} 
can be rewritten in the form
\begin{subequations}
\begin{align}
[\xi , D_+^{\Io}] &= - \hf (\s[\xi] + K[\xi]) D_+^{\Io} - \r[\xi]^{\Io \Jo} D_{+}^{\Jo} ~, \\
[\xi , D_-^{\Iu}] &= - \hf (\s[\xi] - K[\xi]) D_-^{\Iu} - \r[\xi]^{\Iu \Ju} D_{-}^{\Ju} ~,
\end{align}
\end{subequations}
where we have defined the following parameters:
\begin{subequations}
	\label{2.8}
\begin{align}
	\s[\xi] &:= \hf \big ( \partial_{++} \xi^{++} + \partial_{--} \xi^{--} \big)~, \label{2.8a}\\
	K[\xi] &:= \hf \big ( \partial_{++} \xi^{++} - \partial_{--} \xi^{--} \big)~, \label{2.8b}\\
	\r[\xi]^{\Io \Jo} &:= - \frac{\ri}{4} \big[ D_+^{\Io} , D_+^{\Jo} \big ] \xi^{++} ~, \\
	\r[\xi]^{\Iu \Ju} &:= - \frac{\ri}{4} \big[ D_-^{\Iu} , D_-^{\Ju} \big ] \xi^{--} ~.
\end{align}
\end{subequations}
Their $z$-independent components generate scale, Lorentz, $\mathfrak{so}(p)$  
and $\mathfrak{so}(q)$ transformations, respectively.
We point out that the $\mathfrak{so}(p)$ parameter $\r[\xi]^{\Io \Jo} $ is a function of the left variables $\z_L =(x^{++},\q^+)$, while the $\mathfrak{so}(q)$ parameter $\r[\xi]^{\Iu \Ju} $ is a function of the right variables $\z_R =(x^{--},\q^-)$. Further, the former (latter) identically vanishes when $p<2$ ($q<2$).

Given two conformal Killing supervectors $\xi_1$ and $\xi_2$, their commutator is another conformal Killing supervector $\xi_3$
\begin{subequations}
	\label{bracket}
	\begin{align} 
		[\xi_1 , \xi_2] = \xi_3^{a} \partial_{a} + \xi_3^{+\OI} D_{+}^{\OI} + \xi_3^{- \UI} D_{-}^{\UI} = \xi_3~,
	\end{align}
	with the definitions
	\begin{align}
		\xi_3^{++} &= \xi_1^{++} \pa_{++} \xi_2^{++} - \xi_2^{++} \pa_{++} \xi_1^{++} + 2 \ri \xi_1^{+ \OI} \xi_2^{+ \OI}~ \implies~ D_{-}^{\UI} \xi_3^{++} = 0~, \\
		\xi_3^{--} &= \xi_1^{--} \pa_{--} \xi_2^{--} - \xi_2^{--} \pa_{--} \xi_1^{--} + 2 \ri \xi_1^{- \UI} \xi_2^{- \UI} ~ \implies ~ D_{+}^{\OI} \xi_3^{--} = 0~, 
	\end{align}
\end{subequations}
and the spinor parameters are determined by eq. \eqref{2.6}. 
Equations \eqref{2.5} imply that the algebra of conformal Killing supervector fields of $\mathbb{M}^{(2|p,q)}$ is infinite dimensional.
It
may be referred to as 
a $(p,q)$ super Virasoro algebra. Such superalgebras for the $p=q$
case were studied in \cite{Ademollo:1975an, Ramond:1976qq, Gastmans:1987up}.

The superconformal transformation law of a primary tensor superfield $U$ (with suppressed Lorentz, $\sSO(p)$ and $ \sSO(q)$ indices) is
\bea
\d_\x U &=&\Big\{
  \x + \l_U K[\x] 
+\D_U\s[\x]  + \hf \r[\xi]^{\Io \Jo} \mathfrak{L}^{\OI \OJ} 
+ \hf \r[\xi]^{\Iu \Ju}  \mathfrak{R}^{\UI \UJ} \Big\}U~,
\eea
where $\mathfrak{L}^{\OI \OJ} $ and $ \mathfrak{R}^{\UI \UJ} $ are the generators 
of the groups $\sSO(p)$ and $ \sSO(q)$, respectively.\footnote{As usual, we adopt the convention where a factor of $1/2$ is inserted when performing a summations over pairs of antisymmetric indices.} 
The parameters $\l_U$ and $\D_U$ are called the Lorentz weight and the dimension (or Weyl weight) of $U$, respectively. These weights are related if $U$ depends only on $\z_L$ or $\z_R$,
\begin{subequations}
\bea
U&=& U(\z_L) \quad \implies \quad \l_U = \D_U~,\qquad  \mathfrak{R}^{\UI \UJ} U=0~, \\
U&=& U(\z_R) \quad \implies \quad \l_U = -\D_U~, \qquad 
\mathfrak{L}^{\OI \OJ} U=0~.
\eea
\end{subequations}

As pointed out above, 
the algebra of conformal Killing supervector fields of $\mathbb{M}^{(2|p,q)}$ is infinite dimensional. It contains a finite dimensional subalgebra 
which is singled out by the  constraints
\begin{subequations}
	\label{Truncations}
\begin{align}
	p=0:& \quad \pa_{++} \pa_{++} \pa_{++} \xi^{++}=0~,
	\\
	p=1:& \quad \pa_{++} \pa_{++} D_{+} \xi^{++}= 0~,
	\\
	p=2:& \quad \pa_{++} D_{+}^{[\OI} D_{+}^{\OJ]} \xi^{++}= 0~, 
	\\
	p>2:& \quad D_{+}^{[\OI} D_{+}^{\OJ} D_+^{\OK]} \xi^{++}= 0~, 
\end{align}
\end{subequations}
and their counterparts in the right sector.
Physically, these conditions mean that $\xi$ generates those infinitesimal superconformal transformations that belong to the superconformal algebra $\mathfrak{osp}(p|2;\mathbb{R}) \oplus \mathfrak{osp}(q|2;\mathbb{R})$.
This is the Lie algebra of the superconformal group ${\sOSp}_0 (p|2; {\mathbb R} ) \times  {\sOSp}_0 (q|2; {\mathbb R} )$,
which acts on the compactified Minkowski superspace \eqref{B.1}.
It is instructive to check that \eqref{bracket} preserves the conditions \eqref{Truncations}.
We will assume \eqref{Truncations} in what follows.

Employing \eqref{2.5}, it is possible to show that the parameters 
\eqref{2.8a} and \eqref{2.8b} 
satisfy the constraints
\begin{subequations}
	\label{2.9}
\begin{align}
	D_+^{\Io} \s[\xi] &=  D_+^{\Io} K[\xi] \quad \implies \quad \pa_{++} \s[\xi] = \pa_{++} K[\xi]~, \\
	D_-^{\Iu} \s[\xi] &=  - D_-^{\Iu} K[\xi] \quad \implies \quad  \pa_{--} \s[\xi] = - \pa_{--} K[\xi]~.
\end{align}
Next, by using \eqref{2.5} in conjunction with \eqref{Truncations}, one obtains the following constraints on the $R$-symmetry parameters:
\begin{align}
	D_+^{\Io} \r[\xi]^{\Jo \OK} &= 2 \d^{\Io [\Jo} D_+^{\OK]} \s[\xi] \quad \implies \quad \pa_{++} \r^{\OI \OJ}[\xi] = 0 ~,\\
	D_-^{\Iu} \r[\xi]^{\Jo \OK} &= 0 \quad \implies \quad \pa_{--} \r^{\OI \OJ}[\xi] = 0~, \\
	D_+^{\OI} \r[\xi]^{\UJ \UK} &= 0  \quad \implies \quad \pa_{++} \r^{\UI \UJ}[\xi] = 0~, \\
	D_-^{\UI} \r[\xi]^{\UJ \UK} &= 2 \d^{\UI [\UJ} D_-^{\UK]} \s[\xi] \quad \implies \quad \pa_{--} \r^{\UI \UJ}[\xi] = 0~,
\end{align}
\end{subequations}
and on the scaling parameter:
\begin{subequations}
	\label{2.11}
\begin{align}
	D_+^{\OI} D_+^{\OJ} \s[\xi] &= \frac{\ri}{p} \d^{\OI \OJ} \partial_{++} \s[\xi] \quad \implies \quad \partial_{++} D_+^\OI \s[\xi] = 0~, \\
	D_-^{\UI} D_-^{\UJ} \s[\xi] &= \frac{\ri}{q} \d^{\UI \UJ} \partial_{--} \s[\xi] \quad \implies \quad \partial_{--} D_-^\UI \s[\xi] = 0~.
\end{align}
\end{subequations}

The above results mean that we may parametrise the conformal Killing supervector fields obeying \eqref{Truncations} as
\be 
\xi \equiv \xi(\l(P)^{a},\l(Q)^{+\OI}, \l(Q)^{-\UI} ,  \l(M) , \l(\mathbb{D}) ,\l(\mathfrak{L})^{\OI\OJ} , \l(\mathfrak{R})^{\UI\UJ} , \l(K)_a , \l(S)_+^{\OI}, \l(S)_-^{\UI}) ~,
\ee
where we have defined the parameters
\bsubeq
\begin{align} \l(P)^a &:= \xi^{a} |_{z = 0} ~, \qquad \l(Q)^{+ \OI} := \xi^{+\OI}|_{z = 0} ~, \qquad \l(Q)^{- \UI} := \xi^{- \UI}|_{z = 0} ~, \\
	& \qquad \quad ~\, \l(M) := K[\xi]|_{z = 0} ~, \qquad \quad \, \l({\mathbb D}) := \s[\xi]|_{z = 0} ~, \\
	& \qquad \quad  \l(\mathfrak{L})^{\OI \OJ} := \r[\xi]^{\OI \OJ}|_{z = 0} ~, \;\,\, \quad \l(\mathfrak{R})^{\UI \UJ} := \r[\xi]^{\UI \UJ}|_{z = 0} \\
	\l(K)_a &:= \hf \pa_a \s[\xi] |_{z = 0} ~, \quad \l(S)_{+}^{\OI} := \hf D_+^\OI \s[\xi]|_{z = 0} ~, \quad \l(S)_{-}^{\UI} := \hf D_-^\UI \s[\xi]|_{z = 0} ~.
\end{align}
\esubeq
In particular, $\xi$ may be represented as
\be
\xi = \l(X)^{\tilde{A}} X_{\tilde{A}} ~,
\ee
where we have introduced a condensed notation for the superconformal parameters
\begin{subequations}
\begin{align}
	\l(X)^{\tilde{A}} &= (\l(P)^{A}, \l(M) , \l(\mathbb{D}) ,\l(\mathfrak{L})^{\OI\OJ} , \l(\mathfrak{R})^{\UI\UJ} , \l(K)_{A})~, \\
	\l(P)^A &= (\l(P)^{a}, \l(Q)^{+\OI}, \l(Q)^{-\UI}) , \qquad \l(K)_A = (\l(K)_a, \l(S)_{+}^{\OI}, \l(S)_{-}^{\UI}) ~,
\end{align}
\end{subequations}
and for generators of the superconformal algebra
\begin{subequations}
\label{2.18}
\begin{align}
	X_{\tilde{A}} &= (P_A, M , \mathbb{D} ,\mathfrak{L}^{\OI\OJ} , \mathfrak{R}^{\UI\UJ} , K^A)~, \\
	P_A &= (P_{a}, Q^{\OI}_+, Q^{\UI}_{-}) ~, \qquad K^A = (K^{a}, S_{+}^{\OI}, S_{-}^{\UI})~.
\end{align}
\end{subequations}

Making use of the above results allows us to derive the graded commutation relations for the superconformal algebra,
keeping in mind the relation
\bea
[\xi_1,\xi_2] = - \l(X)_2^{\tilde{B}} \l(X)_1^{\tilde{A}} \big[ X_{\tilde{A}}, X_{\tilde{B}} \big \}~.
\eea  
The commutation relations for the conformal algebra are as follows:
\begin{subequations}
	\label{2.20}
	\begin{align}
		[M,P_{\pm\pm}]&= \pm P_{\pm\pm}~, \qquad ~\, [\mathbb{D},P_{\pm\pm}]=P_{\pm \pm}~, \label{2.20a}\\ 
		[M,K^{\pm\pm}]&= \mp K^{\pm\pm}~, \qquad [\mathbb{D},K^{\pm\pm}]=-K^{\pm\pm}~,\\
		&\; [K^{\pm \pm},P_{\pm \pm}]= 2 (\mathbb{D} \pm  M) ~. \label{2.20c}
	\end{align}
The $R$-symmetry generators $\mathfrak{L}^{\OI \OJ}$ and $\mathfrak{R}^{\UI \UJ}$ commute with all the generators of the conformal group. Amongst themselves, they obey the algebra
\begin{align}
	[\mathfrak{L}^{\OI \OJ},\mathfrak{L}^{\OK \OL}]&= 2 \d^{\OK [\OI} \mathfrak{L}^{\OJ] \OL} - 2 \d^{\OL [\OI} \mathfrak{L}^{\OJ] \OK} ~,\\
	[\mathfrak{R}^{\UI \UJ},\mathfrak{R}^{\UK \UL}]&= 2 \d^{\UK [\UI} \mathfrak{R}^{\UJ] \UL} - 2 \d^{\UL [\UI} \mathfrak{R}^{\UJ] \UK} ~.
\end{align}
The superconformal algebra is then obtained by extending the translation generator $P_{a}$ to $P_A$ and the special conformal generator $K^{a}$ to $K^A$. The commutation relations involving the $Q$-supersymmetry generators with the bosonic ones are:
	\begin{align}
	\big[M, Q_+^\OI \big] &= \hf Q_+^\OI ~,\qquad \qquad \quad 
	\big[M, Q_-^\UI \big] = - \hf Q_-^\UI~,\\
	\big[\mathbb{D}, Q_+^\OI \big] &= \hf Q_+^\OI ~, \qquad \qquad \quad \,\,
	\big[\mathbb{D}, Q^\UI_- \big] = \hf Q^\UI_- ~, \\
	\big[\mathfrak{L}^{\OI \OJ}, Q_+^\OK \big] &=  2 \d^{\OK[\OI}Q_+^{\OJ]} ~, \qquad \; \;
	\big[\mathfrak{R}^{\UI \UJ}, Q_-^\UK \big] = 2 \d^{\UK[\UI}Q_-^{\UJ]} ~,  \\
	\big[K^{++}, Q_+^\OI \big] &= \ri S^{+ \OI} ~, \qquad \quad \quad \,
	\big[K^{--}, Q_-^\UI \big] = \ri S^{- \UI} ~.
	\end{align}
The commutation relations involving the $S$-supersymmetry generators 
with the bosonic operators are: 
	\begin{align}
	\big[M, S^{+\OI} \big] &= - \hf S^{+\OI} ~,\qquad \qquad \quad 
	\big[M, S^{-\UI} \big] = \hf S^{-\UI}~,\\
	\big[\mathbb{D}, S^{+\OI} \big] &= - \hf S^{+\OI} ~, \qquad \qquad \quad \,\,
	\big[\mathbb{D}, S^{-\UI} \big] = - \hf S^{-\UI} ~, \\
	\big[\mathfrak{L}^{\OI \OJ}, S^{+\OK} \big] &=  2 \d^{\OK[\OI} S^{+\OJ]} ~, \qquad \;\;\;
	\big[\mathfrak{R}^{\UI \UJ}, S^{-\UK} \big] = 2 \d^{\UK[\UI} S^{-\UJ]} ~,  \\
	\big[S^{+\OI}, P_{++} \big] &= -2\ri Q_{+}^{\OI} ~, \qquad \quad \quad \;
	\big[S^{-\UI}, P_{--} \big] = -2\ri Q_{-}^{\UI} ~.
	\end{align}
Finally, the anti-commutation relations of the fermionic generators are: 
	\begin{align}
	\{Q_+^{\OI} , Q_+^{\OJ} \} &= 2 \ri \d^{\OI \OJ} P_{++}~, \qquad \quad\;\;\, \{Q_-^{\UI} , Q_-^{\UJ} \} = 2 \ri \d^{\UI \UJ} P_{--} ~, \\
	\{ S^{+\OI} , S^{+\OJ} \} &= - 4 \ri \d^{\OI \OJ} K^{++}~, \qquad \{ S^{-\UI} , S^{-\UJ} \} = - 4 \ri \d^{\UI \UJ} K^{--}~,\\
	& \;\; \{ S^{+\OI} , Q_+^{\OJ} \} = 2 \d^{\OI \OJ} (\mathbb{D} + M) - 2 \mathfrak{L}^{\OI \OJ} ~, \\
	& \;\; \{ S^{-\UI} , Q_-^{\UJ} \} = 2 \d^{\UI \UJ} (\mathbb{D} - M) - 2 \mathfrak{R}^{\UI \UJ}  ~.
	\end{align}
\end{subequations}
Note that all remaining (anti-)commutators not listed above vanish identically.


\section{Conformal geometry in two dimensions}\label{section3}

Before turning to the superconformal case, it is instructive to first consider conformal gravity as the gauge theory of the $d=2$ conformal group
$\mathsf{SL} (2 ,{\mathbb R} ) \times \mathsf{SL} (2, {\mathbb R} ) $.
Such a formulation can be extracted from those for the $(1,0)$, $\cN=1$ and $\cN=2$  conformal supergravity theories
\cite{vanNieuwenhuizen:1985an, Uematsu:1984zy, Uematsu:1986de, Hayashi:1986ev, Uematsu:1986aa, Bergshoeff:1985qr, Bergshoeff:1985gc, Schoutens:1986kz}. However, our discussion below has some specific features, since it is targeted at  formulating $(p,q)$ conformal supergravity in the next section.
We will also emphasise those aspects of conformal gravity which 
are unique to two dimensions as compared with the $d>2$ case reviewed in appendix \ref{AppendixA}.

We recall from the previous section that the conformal algebra of $\mathbb{M}^{2}$ is spanned by the operators $X_{\tilde{a}} $ comprising 
the translation ($P_{a}$), Lorentz ($M$), dilatation ($\mathbb{D}$) and special conformal generators ($K^{a}$), which can be grouped into the two disjoint subalgebras spanned by $P_a$ and $X_{\underline a}$:
\begin{align}
X_{\tilde{a}} = (P_{a}, X_{\underline{a}})~, \qquad X_{\underline{a}} = (M, \mathbb{D}, K^{a})~.
\end{align}
Then, the commutation relations \eqref{2.20a}-\eqref{2.20c} may be rewritten as follows
\bsubeq
\begin{align}
	[X_{\underline{a}} , X_{\underline{b}} ] &= -f_{\underline{a} \underline{b}}{}^{\underline{c}} X_{\underline{c}} \ , \\
	[X_{\underline{a}} , P_{{b}} ] &= -f_{\underline{a} { {b}}}{}^{\underline{c}} X_{\underline{c}}
	- f_{\underline{a} { {b}}}{}^{ {c}} P_{ {c}} \label{nonsusymixing}
	~,
\end{align}
\esubeq
where $f_{\underline{a} \underline{b}}{}^{\underline{c}}$, $f_{\underline{a} { {b}}}{}^{\underline{c}}$ and $f_{\underline{a} { {b}}}{}^{ {c}}$ denote the structure coefficients of the conformal algebra.


\subsection{Gauging the conformal algebra}

Let $\mathcal{M}^{2}$ be a two-dimensional curved spacetime parametrised by local coordinates $x^m$.
To gauge the conformal algebra we associate each non-translational generator $X_{\underline{a}}$ with a connection one-form $\o^{\underline{a}} = (\o,b,\mathfrak{f}_a)=\rd x^{m} \o_m{}^{\underline{a}}$ and with $P_a$ a vielbein one-form $e^a = \rd x^m e_m{}^a$, where it is assumed that $e:={\rm det}(e_m{}^a) \neq 0$, hence there exists a unique inverse vielbein $e_a{}^m$
\begin{align}
	e_a{}^m e_m{}^b = \d_a{}^b~, \qquad e_m{}^a e_a{}^n=\d_m{}^n~.
\end{align}
The latter may be used to construct
the vector fields $e_a = e_a{}^m \pa_m $, with 
$\pa_m = \pa /\pa x^m$, which constitute a basis for the tangent space at each
point of $\mathcal{M}^{2}$. It may then be used to express the connection in the vielbein basis
as $\omega^{\underline{a}} =e^b\omega_b{}^{\underline{a}}$, 
where $\omega_b{}^{\underline{a}}=e_b{}^m\omega_m{}^{\underline{a}}$. 
The covariant derivatives $\nabla_a$ then take the form 
\bea
\label{CGNabla}
\nabla_a
&=& e_a  - \o_a{}^{\underline b} X_{\underline b}=
e_a -  \o_a M - b_a \mathbb{D} - \mathfrak{f}_{ab} K^b
~.
\eea
It should be noted that the translation generators $P_a$ do not appear in the above expression.
Instead, we assume that they are replaced by the covariant derivatives $\nabla_a$ and obey the graded commutation relations (c.f. \eqref{nonsusymixing})
\be
[ X_{\underline{a}} , \nabla_b \} = - f_{\underline{a} b}{}^{\underline{c}} X_{\underline{c}} -f_{\underline{a} b}{}^c \nabla_c ~.
\ee

By definition, the gauge group of conformal gravity is generated by local transformations of the form
\begin{subequations}\label{CGtransmations}
	\bea
	\delta_{\mathscr K} \nabla_a &=& [\mathscr{K},\nabla_a] \ , \\
	\mathscr{K} &=& \xi^b \nabla_b +  \L^{\underline{b}} X_{\underline{b}}
	=  \xi^b \nabla_b + K M + \s \mathbb{D} + \L_b K^b ~,
	\eea
\end{subequations}
where  the gauge parameters satisfy natural reality conditions. These gauge transformations act
on a conformal tensor field $\mathcal{U}$ (with its indices suppressed) as 
\bea 
\label{CGMattertfs}
\d_{\mathscr K} \mathcal{U} = {\mathscr K} \mathcal{U} ~.
\eea
Further, we will say that $\mathcal{U}$ is primary if (i) it is annihilated by the special
conformal generators, $K^a \mathcal{U} = 0$; and (ii) it is an eigenvector of $\mathbb D$.  It  will be said to have 
dimension $\D$ and Lorentz weight $\l$ if 
\begin{align}
	\mathbb D \cU = \D \cU~, \qquad M \mathcal{U} = \l \mathcal{U}~. 
\end{align}

The covariant derivatives \eqref{CGNabla} obey the commutation relations
\begin{align}
	\big[\nabla_{++} , \nabla_{--}\big] &= -\mathcal{T}^{a} \nabla_a - \mathcal{R}(X)^{\underline{a}} X_{\underline{a}} = - \cT^{++} \nabla_{++} - \cT^{--} \nabla_{--} - \mathcal{R}(X)^{\underline{a}} X_{\underline{a}} ~,
\end{align}
where the torsion and curvatures take the form
\begin{subequations}
\begin{align}
	\mathcal{T}^{++} &= - \mathscr{C}^{++} + \o^{++} + b^{++}~,\\
	\mathcal{T}^{--} &= - \mathscr{C}^{--} + \o^{--} - b^{--} ~,\\
	\mathcal{R}(M) &= - \hf\mathcal{R} - 2 (\mathfrak{f}_{++,--} + \mathfrak{f}_{--,++})~,\\
	\mathcal{R}(\mathbb{D}) &= - \mathscr{C}^a b_a + e_{++} b_{--} - e_{--} b_{++} + 2(\mathfrak{f}_{++,--} - \mathfrak{f}_{--,++}) ~,\\
	\mathcal{R}(K)_{++} &= - \mathscr{C}^a \mathfrak{f}_{a,++} + e_{++} \mathfrak{f}_{--,++} - e_{--} f_{++,++} - \o_{++} f_{--,++} \non \\
	&\phantom{=}~ - b_{++} f_{--,++} + \o_{--} \mathfrak{f}_{++,++} + b_{--} \mathfrak{f}_{++,++}~,\\
	\mathcal{R}(K)_{--} &= - \mathscr{C}^a \mathfrak{f}_{a,--} + e_{++} \mathfrak{f}_{--,--} - e_{--} f_{++,--} \non \\
	&\phantom{=}~ - b_{++} f_{--,--} - \o_{--} \mathfrak{f}_{++,--} + b_{--} \mathfrak{f}_{++,--} + \o_{++} f_{--,--} ~, \\
	\mathcal{R} &= 2 \mathscr{C}^{a} \o_a -2 e_{++} \o_{--} +  e_{--} \o_{++} ~.
\end{align}
\end{subequations}
Here ${\mathcal R}$ is the scalar curvature constructed from the Lorentz connection $\o_{a}$ and we have introduced the anholonomy coefficients $\mathscr{C}^a$
\begin{align}
	[e_{++},e_{--}] = \mathscr{C}^a e_a = \mathscr{C}^{++} e_{++} + \mathscr{C}^{--} e_{--}~.
\end{align}

In order for this geometry to describe conformal gravity, it is necessary to impose certain covariant constraints. Specifically, we require that the torsion and both Lorentz and dilatation curvatures vanish
\begin{align}
\cT^{a} = 0 ~, \qquad \cR(M) = 0~, \qquad \cR(\mathbb{D}) = 0~.
\end{align}
The constraint $\cT^{a} = 0$ determines the Lorentz connection in terms of the vielbein and dilatation connection
\begin{align}
	\o_{\pm\pm} = \mathscr{C}_{\pm\pm} \pm b_{\pm\pm}~,
\end{align}
while $\cR(M) = \cR(\mathbb{D}) = 0$ fixes several components of the special conformal connection
\begin{subequations}
\label{CGSC}
\begin{align}
	\mathfrak{f}_{++,--} &= -\frac{1}{8} \big ( \mathcal{R} -2 \mathscr{C}^a b_a +2 e_{++} b_{--} -2 e_{--} b_{++} \big) ~, \\
	\mathfrak{f}_{--,++} &= -\frac{1}{8} \big ( \mathcal{R} +2 \mathscr{C}^a b_a -2 e_{++} b_{--} +2 e_{--} b_{++} \big) ~.
\end{align}
\end{subequations}
As a result, the algebra of conformal covariant derivatives takes the form\footnote{The $\cN=(1,0)$ superconformal extension of this geometry is described in appendix \ref{AppendixC}.}
\begin{align}
	\label{3.15}
	\big[\nabla_{++} , \nabla_{--}\big] &= W_{++} K^{++} + W_{--} K^{--} ~, \qquad W_{a} \equiv - \mathcal{R}(K)_{a}~,
\end{align}
where $W_{++}$ and $W_{--}$ have the following conformal properties 
\begin{subequations}
\begin{align}
	K^{a} W_{++} = 0 ~, \qquad \mathbb{D} W_{++} = 3 W_{++} ~, \\
	K^{a} W_{--} = 0 ~, \qquad \mathbb{D} W_{--} = 3 W_{--} ~.
\end{align}
\end{subequations}

Strictly speaking, it is necessary to impose the constraint $W_a = 0$ for this geometry to describe conformal gravity.\footnote{This is in contrast to the situation in $d>3$ dimensions, see appendix \ref{AppendixA}.} Specifically, in the absence of this constraint, there are extra degrees of freedom, in addition to the vielbein, which correspond to the special conformal connections $\mathfrak{f}_{++,++}$ and $\mathfrak{f}_{--,--}$. This will be addressed in further detail below.


\subsection{Degauging to Lorentzian geometry}

According to \eqref{CGtransmations}, under an infinitesimal special superconformal gauge transformation $\mathscr{K} = \Lambda_{b} K^{b}$, the dilatation connection transforms algebraically
\bea
\d_{\mathscr{K}} b_{a} = - 2 \L_{a} ~.
\eea
As a result, we may enforce the gauge $b_{a} = 0$, which completely fixes 
the freedom to perform special superconformal transformations with unconstrained $\L_b$. Hence, the connection $\mathfrak{f}_{ab}$ is not required for the covariance of $\nabla_a$ and it may be separated
\bea
\nabla_{a} &=& \cD_{a} - \mathfrak{f}_{ab} K^{b} ~,
\eea
where the operator $\cD_a$ takes the form
\begin{align}
	\cD_a = e_a -  \o_a M ~.
\end{align}

An important feature of this gauge, which follows from \eqref{CGSC}, is 
\bea
\mathfrak{f}_{++,--} = \mathfrak{f}_{--,++} = - \frac{1}{8} \mathcal{R} ~,
\eea
which, keeping in mind the relation
\bea
\label{CGDegaugedGeom}
[ \cD_{++} , \cD_{--} ] &=& [ \nabla_{++} , \nabla_{--} ] + \big(\cD_{++} \mathfrak{f}_{--,a} - \cD_{--} \mathfrak{f}_{++,a} \big) K^a \non \\
&\phantom{=}& + \mathfrak{f}_{++,a} [ K^{a} , \nabla_{--}] - \mathfrak{f}_{--,a} [ K^{a} , \nabla_{++}]  ~,
\eea
allows one to determine $[ \cD_{++} , \cD_{--} ]$ by a routine computation. One finds
\begin{align}
	[ \cD_{++}, \cD_{--}] = \hf \cR M~.
\end{align}
Additionally, by analysing the special conformal sector of \eqref{CGDegaugedGeom}, we obtain the relations
\begin{align}
	W_{++} = -\frac{1}{8} \cD_{++} \cR -\cD_{--} \mathfrak{f}_{++,++} ~, \qquad
	W_{--} = \frac{1}{8} \cD_{--} \cR + \cD_{++} \mathfrak{f}_{--,--} ~.
	\label{323}
\end{align}
In particular, for vanishing $W_{++}$ ($W_{--}$), we see that $\mathfrak{f}_{++,++}$ ($\mathfrak{f}_{--,--}$) is a non-local function of the vielbein, which is in contrast to the situation in $d>2$ spacetime dimensions (see appendix \ref{AppendixA} for a review).

If no constraint is imposed on the conformal curvature tensors $W_{++}$ and $W_{--}$, then the components $\mathfrak{f}_{++,++} $ and $ \mathfrak{f}_{--,--} $ of the 
special conformal connection remain independent fields in addition to the vielbein. 
Therefore we are forced to impose the constraints 
\bea
W_{++}=0~, \qquad W_{--}=0~,
\eea
in order for the  vielbein to be the only independent field. Then it follows from 
\eqref{323} that $\mathfrak{f}_{++,++} $ and $ \mathfrak{f}_{--,--} $ become non-local functions of the gravitational field. 

Next, it is important to describe the gauge freedom of this geometry, which corresponds to the residual gauge transformations of \eqref{CGtransmations} in the gauge $b_a = 0$. These include local $\mathcal{K}$-transformations of the form
\begin{subequations}\label{CGtransformations}
	\bea
	\delta_{\mathcal K} \cD_A = [\mathcal{K},\cD_A] ~ , \qquad
	\mathcal{K} = \xi^b \cD_b + K M ~,
	\eea
	which act on tensor fields $\mathcal{U}$ (with indices suppressed) as
	\begin{align}
		\d_\cK \cU = \cK \cU ~.
	\end{align}
\end{subequations}

The gauge transformations \eqref{CGtransformations} are not the most general conformal gravity gauge transformations preserving the gauge $b_a=0$.
Specifically, it may be shown that the following transformation also enjoys this property
\begin{align}
	\mathscr{K}(\s) = \s \mathbb{D} + \frac{1}{2} \nabla_b \s K^b \quad \Longrightarrow \quad \d_{\mathscr{K}(\s)} b_a = 0~,
\end{align}
where $\s$ is real but otherwise unconstrained. As a result, it is necessary to consider how this transformation manifests in the degauged geometry
\begin{align}
	\d_{\mathscr{K}(\s)} \nabla_A \equiv \d_\s \nabla_a = \d_\s \cD_a - \d_\s(\mathfrak{f}_{ab} K^b)~.
\end{align}
Employing this relation, we arrive at the transformation laws
\begin{subequations}
	\begin{align}
		\d_\s \cD_{++} &= \s \cD_{++} + \cD_{++} \s M ~, \\
		\d_\s \cD_{--} &= \s \cD_{--} - \cD_{--} \s M ~, \\
		\d_\s \cR &= 2 \s \cR - 4 \cD_{++} \cD_{--} \s ~,
	\end{align}
\end{subequations}
which are exactly the Weyl transformations of spacetime.


\section{Conformal $(p,q)$ superspace}
\label{Section4}

 In section \ref{Section2}, we 
have realised the superconformal algebra 
$\mathfrak{osp}(p|2;\mathbb{R}) \oplus \mathfrak{osp}(q|2;\mathbb{R})$
as the maximal finite-dimensional subalgebra of the $(p,q)$ super Virasoro algebra.
 Now we turn to formulating the corresponding gauge theory. This is known as conformal superspace and is identified with a pair $(\cM^{(2|p,q)}, \nabla)$, where $\mathcal{M}^{(2|p,q)}$ denotes a supermanifold parametrised by local  coordinates $z^M$,
 and $\nabla$ is a covariant derivative associated with the superconformal algebra. The latter is generated by the operators $X_{\tilde A} $, eq.  \eqref{2.18}, which can be grouped into the two disjoint 
 subsets
 $P_A$ and $X_{\underline A}$,
 \bea 
\label{4.1}
X_{\tilde A} = (P_A, X_{\underline{A}} )~, \qquad
X_{\underline{A}} =
( M ,{\mathbb D}, \mathfrak{L}^{\OI\OJ}, \mathfrak{R}^{\UI\UJ} , K^A)~,
\eea 
each of which constitutes a superalgebra:
\bsubeq
\begin{align}
	[P_{ {A}} , P_{ {B}} \} &= -f_{{ {A}} { {B}}}{}^{{ {C}}} P_{ {C}}
	\ , \\
	[X_{\underline{A}} , X_{\underline{B}} \} &= -f_{\underline{A} \underline{B}}{}^{\underline{C}} X_{\underline{C}} \ , \\
	[X_{\underline{A}} , P_{{B}} \} &= -f_{\underline{A} { {B}}}{}^{\underline{C}} X_{\underline{C}}
	- f_{\underline{A} { {B}}}{}^{ {C}} P_{ {C}}
	\ . \label{mixing}
\end{align}
\esubeq
The structure constants above may be determined by comparing with equations \eqref{2.20}.

In order to define covariant derivatives, it is necessary to associate with each non-translational generator $X_{\underline A}$
a connection one-form 
$\Omega^{\underline{A}} = (\O,B,\Phi^{\OI\OJ},\F^{\UI\UJ},\frak{F}_{A})= \rd z^M \Omega_M{}^{\underline{A}}$,  
and with $P_{ {A}}$ a supervielbein one-form
$E^{ {A}} = (E^{a}, E^{+ \OI},E^{-\UI}) = \rd z^{ {M}} E_M{}^A$. 
It is assumed that the supermatrix $E_M{}^A$ is nonsingular, $E:= {\rm Ber} (E_M{}^A) \neq 0$, 
hence there exists a unique inverse supervielbein $E_A{}^M$
\begin{align}
	E_A{}^ME_M{}^B=\d_A{}^B~, \qquad E_M{}^AE_A{}^N=\d_M{}^N~.
\end{align}
The latter may be used to construct
the supervector fields $E_A = E_A{}^M \pa_M $, with 
$\pa_M = \pa /\pa z^M$, which constitute a basis for the tangent space at each
point of $\mathcal{M}^{(2|p,q)}$. The connection may then be expressed in 
the supervielbein basis as $\Omega^{\underline{A}} =E^B\Omega_B{}^{\underline{A}}$, 
where $\Omega_B{}^{\underline{A}}=E_B{}^M\Omega_M{}^{\underline{A}}$. 
The covariant derivatives $\nabla_A= (\nabla_{a}, \nabla_{+}^{\OI}, \nabla_{-}^{\UI})$ then take the form
\bea
\label{3.4}
\nabla_A 
&=& E_A  - \O_A{}^{\underline B} X_{\underline B}=
E_A -  \Omega_A M - B_A \mathbb{D} - \hf \Phi_A^{\OI \OJ} \mathfrak{L}^{\OI \OJ} - \hf \Phi_A^{\UI \UJ} \mathfrak{R}^{\UI \UJ} - \mathfrak{F}_{AB} K^B
~.
\eea
It should be noted that the translation generators $P_A$ do not appear in the above expression.
Instead, we assume that they are replaced by the covariant derivatives $\nabla_A$ and obey the graded commutation relations
\be
[ X_{\underline{B}} , \nabla_A \} = -f_{\underline{B} A}{}^C \nabla_C
- f_{\underline{B} A}{}^{\underline{C}} X_{\underline{C}} ~,
\ee
where the relevant structure constants were defined in equation \eqref{mixing}.

By definition, the gauge group of conformal supergravity  is generated by local transformations of the form
\begin{subequations}\label{SUGRAtransmations}
	\bea
	\delta_{\mathscr K} \nabla_A &=& [\mathscr{K},\nabla_A] \ , \\
	\mathscr{K} &=& \xi^B \nabla_B +  \L^{\underline{B}} X_{\underline{B}}
	=  \xi^B \nabla_B + K M + \s \mathbb{D} + \hf \r^{\OI \OJ} \mathfrak{L}^{\OI \OJ}
	+ \hf \r^{\UI \UJ} \mathfrak{R}^{\UI \UJ}	+ \L_B K^B ~,~~~~
	\eea
\end{subequations}
where  the gauge parameters satisfy natural reality conditions. These gauge transformations act
on a conformal tensor superfield $\mathcal{U}$ (with its indices suppressed) as 
\bea 
\label{3.14}
\d_{\mathscr K} \mathcal{U} = {\mathscr K} \mathcal{U} ~.
\eea
Further, we will say that $\mathcal{U}$ is primary if (i) it is annihilated by the special
conformal generators, $K^A \mathcal{U} = 0$; and (ii) it is an eigenvector of $\mathbb D$. The superfield is said to have 
dimension $\D$ and Lorentz weight $\l$ if 
$\mathbb D \cU = \D \cU$ and  $ M \mathcal{U} = \l \mathcal{U}$. 

The covariant derivatives \eqref{3.4} obey the graded commutation relations
\begin{align}
	\label{3.8}
	\big[\nabla_{A} , \nabla_B\big\} = -\mathcal{T}_{AB}{}^{C} \nabla_C - \mathcal{R}(X)_{AB}{}^{\underline{C}} X_{\underline{C}}~.
\end{align}
In conformal superspace, we impose the requirement that torsion $\mathcal{T}_{AB}{}^{C}$ and curvature tensors $\mathcal{R}(X)_{AB}{}^{\underline{C}}$ differ from their flat counterparts only by terms proportional 
to the conformal curvatures of $\cM^{(2|p,q)}$. Here we will assume that all such superfields vanish\footnote{See appendix \ref{AppendixC} for a construction of conformal $(1,0)$ superspace with non-vanishing curvature.} and thus the only non-vanishing graded commutators are:
\begin{align}
\label{4.9}
\{ \nabla_{+}^{\Io} , \nabla_{+}^{\Jo} \} =  2 \ri \d^{\Io \Jo} \nabla_{++} ~, \qquad  
\{ \nabla_{-}^{\Iu} , \nabla_{-}^{\Ju} \} =  2 \ri \d^{\Iu \Ju} \nabla_{--} ~.
\end{align}

\section{The superspace geometry of $(p,q)$ supergravity}\label{Section5}

According to \eqref{SUGRAtransmations}, under an infinitesimal special superconformal gauge transformation $\mathscr{K} = \Lambda_{B} K^{B}$, the dilatation connection transforms algebraically
\bea
\d_{\mathscr{K}} B_{A} = - 2 \Lambda_{A} ~.
\eea
Hence, we may enforce the gauge $B_{A} = 0$, which completely fixes 
the freedom to perform special superconformal transformations with unconstrained $\L_B$. As a result, the corresponding connection $\mathfrak{F}_{AB}$ is not required for the covariance of $\nabla_A$, and it may be separated
\bea
\nabla_{A} &=& \cD_{A} - \mathfrak{F}_{AB} K^{B} ~. \label{ND}
\eea
Here the degauged covariant derivative $\cD_A$ involves only the Lorentz and $R$-symmetry connections (depending on the choice of $p$ and $q$).
Additionally, the special superconformal connection
$\mathfrak{F}_{AB}$ may be related to the torsion and curvatures of the degauged geometry by analysing the relation
\bea
\label{4.4}
[ \cD_A , \cD_B \}  &=&[ \nabla_{A} , \nabla_{B} \}+ \big(\cD_A \mathfrak{F}_{BC} - (-1)^{AB} \cD_B \mathfrak{F}_{AC} \big) K^C + \mathfrak{F}_{AC} [ K^{C} , \nabla_B \} \non \\
&& - (-1)^{AB} \mathfrak{F}_{BC} [ K^{C} , \nabla_A \} - (-1)^{BC} \mathfrak{F}_{AC} \mathfrak{F}_{BD} [K^D , K^C \} ~.
\eea

We will refer to the superspace geometry described by the covariant derivatives $\cD_A$ as
curved $\sSO(p) \times \sSO(q)$ superspace. 


\subsection{$p,q > 1$ case}

First, we consider the case where $p,q >1$. By a routine calculation, one finds that the degauged connection $\mathfrak{F}_{AB}$ takes the form
\begin{subequations}
	\label{4.5}
	\bea
	\mathfrak{F}_{+,-}^{\OI \phantom{,,} \UJ} & = & - \mathfrak{F}_{-,+}^{\UJ \phantom{,,} \OI}  = S^{\OI \UJ} ~, \quad
	\mathfrak{F}_{+,+}^{\OI \phantom{,,} \OJ} =  - \mathfrak{F}_{+,+}^{\OJ \phantom{,,} \OI}  = X_{++}^{\OI \OJ}~, \quad
	\mathfrak{F}_{-,-}^{\UI \phantom{,,} \UJ} =  - \mathfrak{F}_{-,-}^{\UJ \phantom{,,} \UI}  = X_{--}^{\UI \UJ}~, \quad
	\\
	\mathfrak{F}_{+, --}^{\OI} &=& \mathfrak{F}_{--, +}^{~~~~\phantom{,}\OI}  = \frac{\ri}{q} \cD_-^{\UJ} S^{\OI \UJ} ~, \qquad \qquad \;\;\,\,
	\mathfrak{F}_{-, ++}^{\UI} = \mathfrak{F}_{++, -}^{~~~~\phantom{,}\UI}  = - \frac{\ri}{p} \cD_+^{\OJ} S^{\OJ \UI} ~, 
	\\
	\mathfrak{F}_{+, ++}^{\OI} &=& \mathfrak{F}_{++, +}^{~~~~\phantom{,}\OI}  = - \frac{\ri}{p-1} \cD_+^{\OJ} X_{++}^{\OJ \OI} ~, \qquad
	\mathfrak{F}_{-, --}^{\UI} = \mathfrak{F}_{--, -}^{~~~~\phantom{,}\UI}  = - \frac{\ri}{q-1} \cD_-^{\UJ} X_{--}^{\UJ \UI} ~, 
	\\
	&& \mathfrak{F}_{++, --} = \mathfrak{F}_{--,++} =  \frac{1}{2pq} \big [ \cD_+^\OI , \cD_-^\UJ \big ] S^{\OI \UJ} - \frac{p+q}{pq} S^{\OI \UJ} S^{\OI \UJ}~,
	\\
	&& \qquad \mathfrak{F}_{++, ++} = \frac{1}{p(p-1)} \cD_+^\OI \cD_+^\OJ X_{++}^{\OI \OJ} - \frac{2}{p} X_{++}^{\OI\OJ} X_{++}^{\OI\OJ} ~,
	\\
	&& \qquad \mathfrak{F}_{--, --} = \frac{1}{q(q-1)} \cD_-^\UI \cD_-^\UJ X_{--}^{\UI \UJ} - \frac{2}{q} X_{--}^{\UI\UJ} X_{--}^{\UI\UJ} ~,
	\eea
\end{subequations}
where we have introduced the {\it imaginary} dimension-$1$ torsion tensors $S^{\OI \UJ}$, $X_{++}^{\OI \OJ}$ and $X_{--}^{\UI \UJ}$.
In contrast to conformal gravity, all components of ${\mathfrak F}_{AB}$ are determined in terms of the supergravity multiplet.

The torsion tensors obey the Bianchi identities
\begin{subequations} \label{(p,q)Bianchi}
\begin{align}
	\cD_+^{\OI} S^{\OJ \UK} &= \frac 1 p \d^{\OI \OJ} \cD_+^{\OL} S^{\OL \UK} + \cD_-^\OK X_{++}^{\OI \OJ} ~, \quad 
	\cD_-^{\UI} S^{\OJ \UK} = \frac 1 q \d^{\UI \UK} \cD_-^{\UL} S^{\OJ \UL} - \cD_+^\OJ X_{--}^{\UI \UK} ~, \\
	\cD_+^{\OI} X_{++}^{\OJ \OK} &= \frac{2}{p-1} \d^{\OI [\OJ} \cD_+^{|\OL} X_{++}^{\OL| \OK]} ~, \qquad \quad 
	\cD_-^{\UI} X_{--}^{\UJ \UK} = \frac{2}{q-1} \d^{\UI [\UJ} \cD_-^{|\UL} X_{--}^{\UL| \UK]} ~.
\end{align}
\end{subequations}
Additionally, it may be shown that the algebra obeyed by $\cD_A$ takes the form
\begin{subequations} \label{(p,q)algebra}
	\bea
	\{ \cD_{+}^{\OI}, \cD_{+}^{\OJ} \} &=& 2 \ri \d^{\OI \OJ} \cD_{++} - 4 X_{++}^{\OK (\OI} \mathfrak{L}^{\OJ) \OK} ~, \\
	\{ \cD_{+}^{\OI}, \cD_{-}^{\UJ} \} &=& - 4 S^{\OI \UJ} M + 2 S^{\OK \UJ} \mathfrak{L}^{\OK \OI} - 2 S^{\OI \UK} \mathfrak{R}^{\UK \UJ} ~, \label{(p,q)algebra.b}\\
	\{ \cD_{-}^{\UI}, \cD_{-}^{\UJ} \} &=& 2 \ri \d^{\UI \UJ} \cD_{--} - 4 X_{--}^{\UK (\UI} \mathfrak{R}^{\UJ) \UK} ~, \\
	\big[ \cD_{+}^{\OI} , \cD_{--} \big]
	& = & - 2 \ri S^{\OI \UJ} \cD_{-}^{\UJ} - \frac{4 \ri}{q} \cD_{-}^{\UJ} S^{\OI \UJ} M + \frac{2 \ri}{q} \cD_{-}^{\UK} S^{\OJ \UK} \mathfrak{L}^{\OJ \OI}
	~, \\
	\big[ \cD_{-}^{\UI} , \cD_{++} \big]
	& = & 2 \ri S^{\OJ \UI} \cD_{+}^{\OJ} - \frac{4 \ri}{p} \cD_{+}^{\OJ} S^{\OJ \UI} M - \frac{2 \ri}{p} \cD_{+}^{\OK} S^{\OK \UJ} \mathfrak{R}^{\UJ \UI} ~,
	\\
	\big[ \cD_{+}^{\OI} , \cD_{++} \big]
	& = & - 2 \ri X_{++}^{\OI \OJ} \cD_+^{\OJ} - \frac{2 \ri}{p-1} \cD_+^\OJ X_{++}^{\OJ \OK} \mathfrak{L}^{\OK \OI} ~,
	\\
	\big[ \cD_{-}^{\UI} , \cD_{--} \big]
	& = & - 2 \ri X_{--}^{\UI \UJ} \cD_-^{\UJ} - \frac{2 \ri}{q-1} \cD_-^\UJ X_{--}^{\UJ \UK} \mathfrak{R}^{\UK \UI}  ~,
	\\
	\big[ \cD_{++} , \cD_{--} \big]
	& = & -\frac{2}{p} \cD_+^{\OJ} S^{\OJ \UI} \cD_-^{\UI} - \frac{2}{q} \cD_-^{\UJ} S^{\OI \UJ} \cD_+^{\OI} \non \\
	&& - \frac{2}{pq} \big( \big[\cD_+^{\OI}, \cD_-^{\UJ}\big] - 2(p+q) S^{\OI \UJ} \big) S^{\OI \UJ} M
	~.
	\eea
\end{subequations}

Next, it is important to describe the supergravity gauge freedom of this geometry, which corresponds to the residual gauge transformations of \eqref{SUGRAtransmations} in the gauge $B_A = 0$. These include local $\mathcal{K}$-transformations of the form
\begin{subequations}\label{(p,q)transformations}
	\bea
	\delta_{\mathcal K} \cD_A &=& [\mathcal{K},\cD_A] \ , \\
	\mathcal{K} &=& \xi^B \cD_B + K M +\hf \r^{\OI \OJ} \mathfrak{L}^{\OI \OJ}
	+ \hf \r^{\UI \UJ} \mathfrak{R}^{\UI \UJ} ~,
	\eea
which act on tensor superfields $\mathcal{U}$ (with indices suppressed) as
\begin{align}
	\d_\cK \cU = \cK \cU ~.
\end{align}
\end{subequations}
The gauge transformations \eqref{(p,q)transformations} prove to not be the most general conformal supergravity gauge transformations preserving the gauge $B_A=0$.
Specifically, it may be shown that the following transformation also enjoys this property
\begin{align}
	\mathscr{K}(\s) = \s \mathbb{D} + \frac{1}{2} \nabla_B \s K^B \quad \Longrightarrow \quad \d_{\mathscr{K}(\s)} B_A = 0~,
\end{align}
where $\s$ is real but otherwise unconstrained. 

As a result, it is necessary to consider how this transformation manifests in the degauged geometry
\begin{align}
	\d_{\mathscr{K}(\s)} \nabla_A \equiv \d_\s \nabla_A = \d_\s \cD_A - \d_\s (\mathfrak{F}_{AB} K^B)~.
\end{align}
Employing this relation, we arrive at the transformation laws for $\cD_A$
\begin{subequations}
\begin{align}
	\d_\s \cD_+^\OI &= \hf \s \cD_+^\OI + \cD_+^\OI \s M - \cD_+^\OJ \s \mathfrak{L}^{\OJ \OI} ~, \\
	\d_\s \cD_-^\UI &= \hf \s \cD_-^\UI - \cD_-^\UI \s M - \cD_-^\UJ \s \mathfrak{R}^{\UJ \UI} ~, \\
	\d_\s \cD_{++} &= \s \cD_{++} - \ri \cD_+^\OI \s \cD_+^\OI + \cD_{++} \s M ~, \\
	\d_\s \cD_{--} &= \s \cD_{--} - \ri \cD_-^\UI \s \cD_-^\UI - \cD_{--} \s M ~,
\end{align}
and, by making use of \eqref{4.5}, it may be shown that the torsions transform as follows
\begin{align}
	\d_\s S^{\OI \UJ} &= \s S^{\OI \UJ} + \frac 1 2 \cD_+^\OI \cD_-^\UJ \s ~, \\
	\d_\s X_{++}^{\OI \OJ} &= \s X_{++}^{\OI \OJ} + \frac 1 4 \big[\cD_+^\OI , \cD_+^\OJ \big]  \s ~, \\
	\d_\s X_{--}^{\UI \UJ} &= \s X_{--}^{\UI \UJ} + \frac 1 4 \big[\cD_-^\UI , \cD_-^\UJ \big]  \s ~.
\end{align}
\end{subequations}
These are the super-Weyl transformations of the degauged geometry.

It should be mentioned that, for the special case $\cN= (2,2)$, equivalent superspace geometry was formulated in the works \cite{Howe:1987ba, Grisaru:1994dm,Grisaru:1995dr, Gates:1995du}.\footnote{This follows from the fact that the superconformal groups $\mathsf{OSp}_0 (2|2; {\mathbb R} ) \times  \mathsf{OSp}_0 (2|2; {\mathbb R} )$ and  
 $\sSU(1,1|1) \times  \sSU(1,1|1) $ are isomorphic.} To see this, we first eliminate the superfields $X_{++}^{\overline{I} \overline{J}}$ and $X_{--}^{\underline{I} \underline{J}}$ in \eqref{(p,q)algebra} by redefining the vector covariant derivatives
\begin{align}
\hat{\cD}_{++} = \cD_{++} -  \ri X_{++}^{\overline{I} \overline{J}} \mathfrak{L}^{\overline{I} \overline{J}} ~, \qquad
\hat{\cD}_{--} = \cD_{--} -  \ri X_{--}^{\underline{I} \underline{J}} \mathfrak{R}^{\underline{I} \underline{J}} ~.
\end{align}
The resulting algebra is as follows
\begin{subequations} \label{(2,2)algebra}
\bea
\{ \cD_{+}^{\OI}, \cD_{+}^{\OJ} \} &=& 2 \ri \d^{\OI \OJ} \hat{\cD}_{++} ~,  \qquad
\{ \cD_{-}^{\UI}, \cD_{-}^{\UJ} \} = 2 \ri \d^{\UI \UJ} \hat{\cD}_{--} ~, 
\\
\{ \cD_{+}^{\OI}, \cD_{-}^{\UJ} \} &=& - 4 S^{\OI \UJ} M + 2 S^{\OK \UJ} \mathfrak{L}^{\OK \OI} - 2 S^{\OI \UK} \mathfrak{R}^{\UK \UJ} ~,\\
\big[ \cD_{+}^{\OI} , \hat{\cD}_{++} \big]
& = & 0 ~, \qquad
\big[ \cD_{-}^{\UI} , \hat{\cD}_{--} \big]
= 0  ~,
\\
\big[ \cD_{+}^{\OI} , \hat{\cD}_{--} \big]
& = & - 2 \ri S^{\OI \UJ} \cD_{-}^{\UJ} - 2 \ri \cD_{-}^{\UJ} S^{\OI \UJ} M + \ri \cD_{-}^{\UK} S^{\OJ \UK} \mathfrak{L}^{\OJ \OI} + \ri \cD_-^\UJ S^{\OI \UK} \mathfrak{R}^{\UJ \UK}
~, \\
\big[ \cD_{-}^{\UI} , \hat{\cD}_{++} \big]
& = & 2 \ri S^{\OJ \UI} \cD_{+}^{\OJ} - 2 \ri \cD_{+}^{\OJ} S^{\OJ \UI} M - \ri \cD_+^\OJ S^{\OK \UI} \mathfrak{L}^{\OJ \OK} - \ri \cD_{+}^{\OK} S^{\OK \UJ} \mathfrak{R}^{\UJ \UI} ~,
\\
\big[ \hat{\cD}_{++} , \hat{\cD}_{--} \big]
& = & - \cD_+^{\OJ} S^{\OJ \UI} \cD_-^{\UI} - \cD_-^{\UJ} S^{\OI \UJ} \cD_+^{\OI} - \frac{1}{2} \big( \big[\cD_+^{\OI}, \cD_-^{\UJ}\big] - 8 S^{\OI \UJ} \big) S^{\OI \UJ} M
~, \non \\
&& - \frac 1 4 \cD_-^\UK \cD_+^\OI S^{\OJ \UK} \mathfrak{L}^{\OI \OJ} - \frac 1 4 \cD_+^\OK \cD_-^\UI S^{\OK \UJ} \mathfrak{R}^{\UI \UJ}~.
\eea
\end{subequations}
It should be emphasised that the resulting geometry is described solely in terms of $S^{\OI \UJ}$.
Then, to relate the geometry of \cite{Howe:1987ba, Grisaru:1994dm, Grisaru:1995dr, Gates:1995du} to ours it is necessary to express \eqref{(2,2)algebra} in a complex basis of spinor covariant derivatives
\begin{subequations}
\label{ComplexBasis}
\begin{align}
\cD_+ := \frac{1}{\sqrt 2} (\cD_+^{\overline 1} - \ri \cD_+^{\overline 2}) ~, \qquad \bar{\cD}_+ = -\frac{1}{\sqrt 2} (\cD_+^{\overline 1} + \ri \cD_+^{\overline 2}) ~, \\
\cD_- := \frac{1}{\sqrt 2} (\cD_-^{\underline 1} - \ri \cD_-^{\underline 2}) ~, \qquad \bar{\cD}_- = -\frac{1}{\sqrt 2} (\cD_-^{\underline 1} + \ri \cD_-^{\underline 2})
~.
\end{align}
\end{subequations}
We omit further technical details regarding this procedure, which will appear in a future work. 

To the best of our knowledge, for $p,q> 2$ our superspace geometry described by 
the equations \eqref{(p,q)Bianchi} and \eqref{(p,q)algebra} has not appeared in the literature. In particular, in the  $\cN=(4,4)$ case, the above $\sSO(4) \times \sSO(4)$ superspace geometry differs from the one proposed in \cite{TM} by the choice of structure group, see section \ref{Section6} for the discussion of this formulation.

As follows from \eqref{4.5}, some expressions are ill-defined if at least one of the parameters $p$ and $q$ takes values $0$ or $1$. Each of these cases should be studied separately, 
which is done in the remainder of this section.


\subsection{$p > 1,~ q = 1$ case}

Next, let us consider the case where $p>1,~q=1$. For convenience, we will unambiguously remove bars over left isovector indices, e.g. $\OI \equiv I$. By a routine calculation, one obtains the following components of the degauged connection $\mathfrak{F}_{AB}$
\begin{subequations}
	\label{(p,1)conns}
	\bea
	\mathfrak{F}_{+,-}^{I \phantom{,,}} & = & - \mathfrak{F}_{-,+}^{\phantom{,,} I}  = S^{I} ~, \quad
	\mathfrak{F}_{+,+}^{I \phantom{,,} J} =  - \mathfrak{F}_{+,+}^{J \phantom{,,} I}  = X_{++}^{I J}~, \quad
	\mathfrak{F}_{-,-} = 0~, \quad
	\\
	\mathfrak{F}_{+, --}^{I} &=& \mathfrak{F}_{--, +}^{~~~~\phantom{,}I}  = \ri \cD_- S^{I} ~, \qquad 
	\mathfrak{F}_{-, ++} = \mathfrak{F}_{++, -}  = - \frac{\ri}{p} \cD_+^{I} S^{I} ~, 
	\\
	\mathfrak{F}_{+, ++}^{I} &=& \mathfrak{F}_{++, +}^{~~~~\phantom{,}I}  = - \frac{\ri}{p-1} \cD_+^{J} X_{++}^{J I} ~, 
	\\
	\mathfrak{F}_{++, --} &=& \mathfrak{F}_{--,++} =  \frac{1}{2p} \big [ \cD_+^I , \cD_- \big ] S^{I} - \frac{p+1}{p} S^{I} S^{I}~,
	\\
	\mathfrak{F}_{++, ++} &=& \frac{1}{p(p-1)} \cD_+^I \cD_+^J X_{++}^{I J} - \frac{2}{p} X_{++}^{IJ} X_{++}^{IJ} ~,
	\eea
where we have introduced the dimension-$1$ torsions $S^{I}$ and $X_{++}^{I J}$.
The remaining components of $\mathfrak{F}_{AB}$ do not play a role in the degauged geometry, though they satisfy
\begin{align}
\mathfrak{F}_{-, --} = \mathfrak{F}_{--, -}~,
\quad
\cD_+^I \mathfrak{F}_{-,--} =\cD_{--} S^I~,
\quad
\mathfrak{F}_{--, --} = - \ri \cD_{-} \mathfrak{F}_{-,--}~.
\end{align}
\end{subequations}
These equations imply that $\mathfrak{F}_{-,--}$ is a non-local function of the supergravity multiplet.

The superfields $S^I$ and $X_{++}^{IJ}$ obey the Bianchi identities
\begin{align}
	\cD_+^{I} S^{J} = \frac 1 p \d^{I J} \cD_+^{K} S^{K} + \cD_- X_{++}^{I J} ~, \qquad 
	\cD_+^{I} X_{++}^{J K} = \frac{2}{p-1} \d^{I [J} \cD_+^{|L} X_{++}^{L| K]} ~.
\end{align}
Additionally, it may be shown that the algebra obeyed by $\cD_A$ takes the form
\begin{subequations} \label{(p,1)algebra}
	\bea
	\{ \cD_{+}^{I}, \cD_{+}^{J} \} &=& 2 \ri \d^{I J} \cD_{++} - 4 X_{++}^{K (I} \mathfrak{L}^{J) K} ~, \\
	\{ \cD_{+}^{I}, \cD_{-} \} &=& - 4 S^{I } M + 2 S^{J} \mathfrak{L}^{J I} ~, \\
	\{ \cD_{-}, \cD_{-} \} &=& 2 \ri  \cD_{--} ~, \\
	\big[ \cD_{+}^{I} , \cD_{--} \big]
	& = & - 2 \ri S^{I} \cD_{-} - 4 \ri \cD_{-} S^{I} M + 2 \ri \cD_{-} S^{J} \mathfrak{L}^{J I}
	~, \\
	\big[ \cD_{-} , \cD_{++} \big]
	& = & 2 \ri S^{I} \cD_{+}^{I} - \frac{4 \ri}{p} \cD_{+}^{I} S^{I} M ~,
	\\
	\big[ \cD_{+}^{I} , \cD_{++} \big]
	& = & - 2 \ri X_{++}^{I J} \cD_+^{J} - \frac{2 \ri}{p-1} \cD_+^J X_{++}^{J K} \mathfrak{L}^{K I} ~,
	\\
	\big[ \cD_{-} , \cD_{--} \big] &=& 0 ~, \\
	\big[ \cD_{++} , \cD_{--} \big]
	& = & -\frac{2}{p} \cD_+^{I} S^{I} \cD_- - 2 \cD_- S^{I} \cD_+^{I} \non \\
	&& - \frac{2}{p} \big( \big[\cD_+^{I}, \cD_-\big] - 2(p+1) S^{I} \big) S^{I} M
	~.
	\eea
\end{subequations}

The supergravity gauge freedom of the $(p,1)$ geometry \eqref{(p,1)algebra} corresponds to the residual gauge transformations of \eqref{SUGRAtransmations} in the gauge $B_A = 0$. In particular, they include the local $\mathcal{K}$-transformations
\begin{subequations}\label{(p,1)transformations}
	\bea
	\delta_{\mathcal K} \cD_A &=& [\mathcal{K},\cD_A] \ , \\
	\mathcal{K} &=& \xi^B \cD_B + K M + \hf \r^{I J} \mathfrak{L}^{I J}~,
	\eea
which act on tensor superfields $\mathcal{U}$ (with indices suppressed) as $\d_\cK \cU = \cK \cU$.
\end{subequations}
Transformations \eqref{(p,1)transformations} prove to not be the most general conformal supergravity gauge transformations preserving the gauge $B_A=0$.
Specifically, the following transformation also enjoys this property
\begin{align}
	\mathscr{K}(\s) = \s \mathbb{D} + \frac{1}{2} \nabla_B \s K^B \quad \Longrightarrow \quad \d_{\mathscr{K}(\s)} B_A = 0~,
\end{align}
where $\s$ is real but otherwise unconstrained. 

As a result, it is necessary to consider how this transformation manifests in the degauged geometry
\begin{align}
	\d_{\mathscr{K}(\s)} \nabla_A \equiv \d_\s \nabla_A = \d_\s \cD_A - \d_\s (\mathfrak{F}_{AB} K^B)~.
\end{align}
Employing this relation, we arrive at the transformation laws for $\cD_A$
\begin{subequations}
	\begin{align}
		\d_\s \cD_+^I &= \hf \s \cD_+^I + \cD_+^I \s M - \cD_+^J \s \mathfrak{L}^{J I} ~, \\
		\d_\s \cD_- &= \hf \s \cD_- - \cD_-\s M~, \\
		\d_\s \cD_{++} &= \s \cD_{++} - \ri \cD_+^I \s \cD_+^I + \cD_{++} \s M ~, \\
		\d_\s \cD_{--} &= \s \cD_{--} - \ri \cD_- \s \cD_- - \cD_{--} \s M ~,
	\end{align}
	and, by making use of \eqref{4.5}, it may be shown that the torsions transform as follows
	\begin{align}
		\d_\s S^{I} &= \s S^{I} + \frac 1 2 \cD_+^I \cD_- \s ~, \\
		\d_\s X_{++}^{I J} &= \s X_{++}^{I J} + \frac 1 4 \big [ \cD_+^I , \cD_+^J \big ]  \s ~.
	\end{align}
\end{subequations}
These are the super-Weyl transformations of the degauged geometry.

It may be shown that, for the special case $p=2$, the superfield $X_{++}^{IJ}$ can be eliminated by performing the redefinition
\begin{align}
	\label{5.17}
	\hat{\cD}_{++} = \cD_{++} -  \ri X_{++}^{\overline{I} \overline{J}} \mathfrak{L}^{\overline{I} \overline{J}} ~.
\end{align}
The resulting algebra then takes the form
\begin{subequations} \label{(2,1)algebra}
	\bea
	\{ \cD_{+}^{I}, \cD_{+}^{J} \} &=& 2 \ri \d^{I J} \hat{\cD}_{++} ~, \qquad \{ \cD_{-}, \cD_{-} \} = 2 \ri  \cD_{--} ~, \\
	\{ \cD_{+}^{I}, \cD_{-} \} &=& - 4 S^{I } M + 2 S^{J} \mathfrak{L}^{J I} ~, \\
	\big[ \cD_{+}^{I} , \cD_{--} \big]
	& = & - 2 \ri S^{I} \cD_{-} - 4 \ri \cD_{-} S^{I} M + 2 \ri \cD_{-} S^{J} \mathfrak{L}^{J I}
	~, \\
	\big[ \cD_{-} , \hat{\cD}_{++} \big]
	& = & 2 \ri S^{I} \cD_{+}^{I} - 2 \ri \cD_{+}^{I} S^{I} M ~,
	\\
	\big[ \cD_{+}^{I} , \hat{\cD}_{++} \big]
	& = & 0 ~, \qquad \big[ \cD_{-} , \cD_{--} \big]
	=  0 ~,
	\\
	\big[ \hat{\cD}_{++} , \cD_{--} \big]
	& = & - \cD_+^{I} S^{I} \cD_- - 2 \cD_- S^{I} \cD_+^{I} - \big( \big[\cD_+^{I}, \cD_-\big] - 6 S^{I} \big) S^{I} M \non \\
	&&- \frac 1 4 \cD_- \cD_+^I S^{J} \mathfrak{L}^{IJ}
	~.
	\eea
\end{subequations}
Hence, the resulting geometry is described solely in terms of $S^{I}$.

\subsection{$p > 1,~ q = 0$ case}
Now, we consider the case $p >1,~ q=0$. As in the previous subsection, we remove bars over left isovector indices; $\OI \equiv I$. By a routine calculation, we readily obtain the following components of the degauged connection $\mathfrak{F}_{AB}$
\begin{subequations}
	\bea
	\mathfrak{F}_{+,+}^{I \phantom{,,} J} &=& - \mathfrak{F}_{+,+}^{J \phantom{,,} I}  = X_{++}^{I J}~,  \\
	\mathfrak{F}_{+, --}^{I} &=& \mathfrak{F}_{--, +}^{~~~~\phantom{,}I}  = G_{-}^{I} ~, \qquad \qquad \phantom{-} \,
	\mathfrak{F}_{+, ++}^{I} = \mathfrak{F}_{++, +}^{~~~~\phantom{,}I}  = - \frac{\ri}{p-1} \cD_+^{J} X_{++}^{J I} ~, \\
	\mathfrak{F}_{++, --} &=& \mathfrak{F}_{--,++} =  -\frac \ri p \cD_+^I G_-^I~, \quad 
	\mathfrak{F}_{++, ++} = \frac{1}{p(p-1)} \cD_+^I \cD_+^J X_{++}^{I J} - \frac{2}{p} X_{++}^{I J} X_{++}^{I J} ~, \qquad
	\eea
while $\mathfrak{F}_{--,--}$ only appears in \eqref{4.4} through the differential constraint
	\begin{align}
		\label{4.12d}
		\cD_+^I \mathfrak{F}_{--,--} = \cD_{--} G_{-}^{I}~.
	\end{align}
\end{subequations}
In the above equations we have introduced the torsions $X_{++}^{I J}$ and $G_{-}^I$, which obey the Bianchi identities
\begin{align}
	\label{5.24}
	\cD_+^{I} X_{++}^{J K} = \frac{2}{p-1} \d^{I [J} \cD_+^{|L} X_{++}^{L| K]} ~, \qquad
	\cD_+^{I} G_-^{J} = \frac{1}{p} \d^{I J} \cD_+^{K} G_-^{K} + \cD_{--} X_{++}^{I J} ~.
\end{align}

The latter constraint in \eqref{5.24} may be solved by expressing $G_-^I$
in terms of the dimension-$1$ superfield $G_{--}$
\begin{align}
	G_-^I = \cD_+^I G_{--} \quad \implies \quad \cD_+^{[I} \cD_+^{J]} G_{--} = \cD_{--} X_{++}^{I J}~.
\end{align}
However, we will not make use of this solution in what follows.

Making use of these results, we find that the algebra obeyed by $\cD_A$ takes the form
\begin{subequations} \label{(p,0)algebra}
	\bea
	\{ \cD_{+}^{I}, \cD_{+}^{J} \} &=& 2 \ri \d^{I J} \cD_{++} - 4 X_{++}^{K (I} \mathfrak{L}^{J) K} ~, \\
	\big[ \cD_{+}^{I} , \cD_{--} \big]
	& = & -4 G_-^I M + 2 G_-^J \mathfrak{L}^{J I}
	~, \\
	\big[ \cD_{+}^{I} , \cD_{++} \big]
	& = & - 2 \ri X_{++}^{I J} \cD_+^{J} - \frac{2 \ri}{p-1} \cD_+^J X_{++}^{J K} \mathfrak{L}^{K I} ~,
	\\
	\big[ \cD_{++} , \cD_{--} \big]
	& = & 2 \ri G_-^I \cD_+^I - \frac{4 \ri}{p} \cD_{+}^I G_{-}^I M ~.
	\eea
\end{subequations}
It should be noted that $\cN=(p,0)$ superspace geometries, with $p \geq 2$, were discussed in \cite{EO}, where the structure was chosen to be the Lorentz group,
$\sSO(1,1)$,
though the $p=2$ case appeared earlier \cite{BMG}.\footnote{The structure group 
for $(2,0) $ supergravity was enlarged from $\sSO(1,1)$ to 
$\sSO(1,1) \times \sU(1)$  in 
\cite{Govindarajan:1991sx}.}
The emergence of the $(2,0)$ geometry proposed in \cite{BMG} in our framework will be discussed below.

The supergravity gauge transformations of the $(p,0)$ geometry \eqref{(p,0)algebra} may be obtained from \eqref{SUGRAtransmations} after imposing $B_A = 0$. They include the local $\mathcal{K}$-transformations
\begin{subequations}\label{(p,0)transformations}
	\bea
	\delta_{\mathcal K} \cD_A &=& [\mathcal{K},\cD_A] \ , \\
	\mathcal{K} &=& \xi^B \cD_B + K M + \hf \r^{I J} \mathfrak{L}^{I J} ~,
	\eea
which act on tensor superfields $\mathcal{U}$ (with indices suppressed) as $\d_\cK \cU = \cK \cU$.
\end{subequations}
In addition to \eqref{(p,0)transformations}, the following transformation also preserves the gauge $B_A=0$
\begin{align}
	\label{(p,0)DGTF}
	\mathscr{K}(\s) = \s \mathbb{D} + \frac{1}{2} \nabla_B \s K^B \quad \Longrightarrow \quad \d_{\mathscr{K}(\s)} B_A = 0~,
\end{align}
where $\s$ is real but otherwise unconstrained. 

It is then necessary to consider how this transformation manifests in the degauged geometry
\begin{align}
	\d_{\mathscr{K}(\s)} \nabla_A \equiv \d_\s \nabla_A = \d_\s \cD_A - \d_\s (\mathfrak{F}_{AB} K^B) ~.
\end{align}
Employing this relation, we arrive at the transformation laws for $\cD_A$
\begin{subequations}
	\label{(p,0)SW}
	\begin{align}
		\d_\s \cD_+^I &= \hf \s \cD_+^I + \cD_+^I \s M - \cD_+^J \s \mathfrak{L}^{J I} ~, \\
		\d_\s \cD_{++} &= \s \cD_{++} - \ri \cD_+^I \s \cD_+^I + \cD_{++} \s M ~, \\
		\d_\s \cD_{--} &= \s \cD_{--} - \cD_{--} \s M ~,
	\end{align}
	and, by making use of \eqref{4.5}, it may be shown that the torsions transform as follows
	\begin{align}
		\d_\s X_{++}^{I J} &= \s X_{++}^{I J} + \frac 1 4 \big [ \cD_+^I , \cD_+^J \big ]  \s ~, \\
		\d_\s G_{-}^I &= \frac{3}2 \s G_{-}^I + \frac 1 2 \cD_+^I \cD_{--} \s ~.
	\end{align}
\end{subequations}
These are exactly the super-Weyl transformations of the degauged geometry.

We emphasise that in special case $p=2$, the superfield $X_{++}^{IJ}$ can be eliminated by performing the redefinition \eqref{5.17}. Additionally, it is useful to work in a complex basis of spinor covariant derivatives, where $\cD_+^I$ is replaced by $\cD_+ = \frac{1}{\sqrt 2} (\cD_+^{1} - \ri \cD_+^{2})$ and its conjugate $\bar \cD_+=- \frac{1}{\sqrt 2} (\cD_+^{1} + \ri \cD_+^{2})$.
The resulting algebra of covariant derivatives is
\begin{subequations} \label{(2,0)algebra}
	\bea
	\{ \cD_{+}, \cD_{+} \} &=& 0 ~, \qquad \{ \cD_{+}, \bar{\cD}_{+} \} = 2 \ri \hat{\cD}_{++} ~, \\
	\big[ \cD_{+} , \cD_{--} \big]
	& = & -4 \bar{G}_- M - 2 \ri \bar{G}_- \mathfrak{L}^{12}
	~, \\
	\big[ \cD_{+}, \hat{\cD}_{++} \big]
	& = &0 ~,
	\\
	\big[ \hat{\cD}_{++} , \cD_{--} \big]
	& = & 2 \ri G_- \cD_+ + 2 \ri \bar{G}_- \bar{\cD}_+ + 2 \ri (\cD_+ {G}_- + \bar{\cD}_+ \bar{G}_-) M \non \\
	&& + (\cD_+ {G}_- - \bar{\cD}_+ \bar{G}_-) \mathfrak{L}^{12}~.
	\eea
\end{subequations}
We see that this supergeometry is described solely in terms of the complex superfield 
\begin{align}
	\label{5.33}
	G_{-} := -\frac{1}{\sqrt 2} (G_-^{1} + \ri G_-^{2}) 
\end{align}
and its conjugate $\bar G_-$
It follows from \eqref{5.24} that it satisfies the chirality constraint
\begin{align}
	\bar{\cD}_{+} G_- = 0~.
\end{align}

The $\sSO(2)$ gauge freedom enjoyed by the geometry \eqref{(2,0)algebra} may be 
used to gauge away its corresponding spinor $\sSO(2)$ connection $\Phi_+^{I,\,JK } = -\Phi_+^{I,\,JK } $. This can be seen   
by coupling the conformal supergravity multiplet to a conformal primary compensator $\phi$ obeying the chirality condition $\bar{\cD}_+ \phi = 0$. Choosing its weight to be $\D_\f =1$,  it transforms under super-Weyl \eqref{(p,0)SW} and local $\sSO(2)$ transformations \eqref{(p,0)DGTF} as follows:
\begin{align}
	\d \phi = (\s - \ri \r^{12} ) \phi ~,
\end{align}
hence this freedom may be used to impose the gauge $\phi = 1$. Associated with this gauge are the consistency conditions
\begin{align}
	\Phi_+^{I,12} = 0 ~, \qquad 
	\ri \bar{\cD}_+ \G_{--} = G_- ~, \qquad
	\G_{--} = \bar{\G}_{--}:= \hf \Phi_{--}^{\phantom{--}12} ~.
\end{align}
As the $\sSO(2)$ connection is now auxiliary, it should be separated by performing the redefinition
\begin{align}
	\hat{\cD}_{--} = \cD_{--} + 2 \G_{--} \mathfrak{L}^{12}~.
\end{align}
The resulting algebra of covariant derivatives is
\begin{subequations} \label{(2,0)algebraDG}
	\bea
	\{ \cD_{+}, \cD_{+} \} &=& 0 ~, \qquad \{ \cD_{+}, \bar{\cD}_{+} \} = 2 \ri \hat{\cD}_{++} ~, \\
	\big[ \cD_{+} , \hat{\cD}_{--} \big]
	& = & -2\ri\G_{--} \cD_+ -4 \ri \bar{\cD}_+ \G_{--} M ~, \\
	\big[ \cD_{+}, \hat{\cD}_{++} \big]
	& = &0 ~,
	\\
	\big[ \hat{\cD}_{++} , \hat{\cD}_{--} \big]
	& = & -2 \bar{\cD}_+ \G_{--} \cD_+ +2 {\cD}_+ \G_{--} \bar{\cD}_+ - 2 [\cD_+, \bar{\cD}_+] \G_{--} M ~,
	\eea
\end{subequations}
which coincides with the one appearing in \cite{BMG}. 

\subsection{$p = q = 1$ case}

Next, we fix $p=q=1$. By a routine calculation, we obtain the following components of the degauged special conformal connection
\begin{subequations}
	\bea
	\mathfrak{F}_{+,+} &=& 0 ~, \qquad \mathfrak{F}_{-,-} = 0 ~, \qquad \mathfrak{F}_{+,-} = - \mathfrak{F}_{-,+} = S  \\
	\mathfrak{F}_{+, --} &=& \mathfrak{F}_{--, +}  = \ri \cD_- S ~, \qquad
	\mathfrak{F}_{-, ++} = \mathfrak{F}_{++, -}  = - \ri \cD_+ S ~, \\
	\mathfrak{F}_{++,++} &=& - \ri \cD_+ \mathfrak{F}_{+,++} ~, \qquad \mathfrak{F}_{--,--} = -\ri \cD_- \mathfrak{F}_{-,--}~, \\
	\mathfrak{F}_{++, --} &=& \mathfrak{F}_{--,++} =  \hf [\cD_+, \cD_-]S - 2 S^2 ~,
	\eea
where we have introduced the real scalar $S$. The remaining components of $\mathfrak{F}_{AB}$ do not play a role in the degauged geometry, though they satisfy the constraints
\begin{align}
	\mathfrak{F}_{+,++} &= \mathfrak{F}_{++,+} ~, \qquad \mathfrak{F}_{-,--} = \mathfrak{F}_{--,-}~, \\
	\cD_- \mathfrak{F}_{+,++} &= - \cD_{++} S ~, \qquad \cD_+ \mathfrak{F}_{-,--} =\cD_{--} S ~.
\end{align}
\end{subequations}
It follows that $\mathfrak{F}_{+,++}$ and $\mathfrak{F}_{-,--}$ are non-local functions of the supergravity multiplet.

It follows from the above results that the algebra obeyed by $\cD_A$ takes the form
\begin{subequations}
	\label{(1,1)algebra}
	\bea
	\{ \cD_{+}, \cD_{+} \} &=& 2 \ri \cD_{++}~, \qquad \{ \cD_{-}, \cD_{-} \} = 2 \ri \cD_{--} ~, \\
	&& \{ \cD_{+}, \cD_{-} \} = - 4 S M~, \\
	\big[ \cD_{+} , \cD_{--} \big]
	& = & - 2 \ri S \cD_- - 4 \ri (\cD_- S) M ~, \\
	\big[ \cD_{-} , \cD_{++} \big]
	& = &  2 \ri S \cD_+ - 4 \ri (\cD_+ S) M ~,
	\\
	\big[ \cD_{++} , \cD_{--} \big]
	& = & -2(\cD_+ S) \cD_- - 2 (\cD_- S) \cD_+ \non \\
	& \phantom{=} &- 2 \big( [\cD_+ , \cD_-] - 4 S \big)S M  ~.
	\eea
\end{subequations}

The supergravity gauge freedom of this geometry may be obtained from the conformal superspace transformations \eqref{SUGRAtransmations} after restricting to the gauge $B_A = 0$. These include local $\mathcal{K}$-transformations of the form
\begin{subequations}\label{(1,1)transformations}
	\bea
	\delta_{\mathcal K} \cD_A &=& [\mathcal{K},\cD_A] ~, \qquad
	\mathcal{K} = \xi^B \cD_B + K M~,
	\eea
	which act on tensor superfields $\mathcal{U}$ (with indices suppressed) as $\d_\cK \cU = \cK \cU$.
\end{subequations}
In addition to \eqref{(1,1)transformations}, it may be shown that the following also preserves the gauge $B_A=0$
\begin{align}
	\mathscr{K}(\s) = \s \mathbb{D} + \frac{1}{2} \nabla_B \s K^B \quad \Longrightarrow \quad \d_{\mathscr{K}(\s)} B_A = 0~,
\end{align}
where $\s$ is real but otherwise unconstrained. 

As a result, it is necessary to consider how this transformation manifests in the degauged geometry
\begin{align}
	\d_{\mathscr{K}(\s)} \nabla_A \equiv \d_\s \nabla_A = \d_\s \cD_A - \d_\s (\mathfrak{F}_{AB} K^B)~.
\end{align}
Employing this relation, we arrive at the transformation laws for $\cD_A$ and $S$
\begin{subequations}
	\begin{align}
		\d_\s \cD_+ &= \hf \s \cD_+ + \cD_+\s M  ~, \\
		\d_\s \cD_- &= \hf \s \cD_- - \cD_-\s M  ~, \\
		\d_\s \cD_{++} &= \s \cD_{++}  - \ri \cD_+ \s \cD_+ + \cD_{++} \s M ~, \\
		\d_\s \cD_{--} &= \s \cD_{--} - \ri \cD_- \s \cD_- - \cD_{--} \s M ~, \\
		\d_\s S &= \s S + \frac 1 2 \cD_+  \cD_- \s~.
	\end{align}
\end{subequations}
These are exactly the super-Weyl transformations of the degauged geometry.

The superspace geometry of $\cN=1$ supergravity described above was originally constructed in \cite{Howe1979,BG}, see also \cite{Martinec,GN,RvanNZ}.


\subsection{$p = 1,~ q = 0$ case} \label{Section5.5}

Finally, let us consider the case $p=1,~q=0$. 
A routine computation leads to the degauged special conformal connections
\begin{subequations}
	\bea
	\label{5.34a}
	\mathfrak{F}_{+,+} &=& 0 ~, \qquad \mathfrak{F}_{+,--} = \mathfrak{F}_{--,+} = G_{-} ~, \qquad \mathfrak{F}_{++,--} = \mathfrak{F}_{--,++} = - \ri \cD_+ G_-~.
	\eea
	where we have introduced the spinor $G_-$. The remaining components of $\mathfrak{F}_{AB}$ do not play a role in the degauged geometry, though they satisfy the constraints
	\begin{align}
		\label{5.34b}
		\mathfrak{F}_{+,++} &= \mathfrak{F}_{++,+} ~, \qquad \mathfrak{F}_{++,++} = - \ri \cD_+ \mathfrak{F}_{+,++}~, \\
		\label{5.34c}
		\cD_{++} G_- &= \cD_{--} \mathfrak{F}_{+,++} ~, \qquad \cD_{--} G_{-} = \cD_+ \mathfrak{F}_{--,--}~.
	\end{align}
\end{subequations}
It is clear that $\mathfrak{F}_{+,++}$ and $\mathfrak{F}_{--,--}$ are non-local functions of the supergravity multiplet.

It immediately follows that the algebra obeyed by $\cD_A$ takes the form
\begin{subequations}
	\label{(1,0)algebra}
	\bea
	\{ \cD_{+}, \cD_{+} \} &=& 2 \ri \cD_{++}~,\\
	\big[ \cD_+, \cD_{--} \big ] &=& - 4 G_- M ~, \quad \big[\cD_{+}, \cD_{++} \big] = 0 \\
	\big[ \cD_{++} , \cD_{--} \big] &=& 2 \ri G_{-} \cD_{+} + 4 \ri (\cD_+ G_-) M  ~.
	\eea
\end{subequations}

As described in the previous subsections, the supergravity gauge transformations of this geometry correspond to \eqref{SUGRAtransmations} in the gauge $B_A = 0$. They include local $\mathcal{K}$-transformations of the form
\begin{subequations}\label{(1,0)transformations}
	\bea
	\delta_{\mathcal K} \cD_A &=& [\mathcal{K},\cD_A] ~, \qquad
	\mathcal{K} = \xi^B \cD_B + \o M~,
	\eea
	which acts on tensor superfields $\mathcal{U}$ (with indices suppressed) as	$\d_\cK \cU = \cK \cU$.
\end{subequations}
One may also show that the following transformation also preserves the $B_A=0$ gauge
\begin{align}
	\mathscr{K}(\s) = \s \mathbb{D} + \frac{1}{2} \nabla_B \s K^B \quad \Longrightarrow \quad \d_{\mathscr{K}(\s)} B_A = 0~,
\end{align}
where $\s$ is real but otherwise unconstrained. 

As a result, it is necessary to consider how this transformation manifests in the degauged geometry
\begin{align}
	\d_{\mathscr{K}(\s)} \nabla_A \equiv \d_\s \nabla_A = \d_\s \cD_A - \d_\s (\mathfrak{F}_{AB} K^B)~.
\end{align}
Employing this relation, we arrive at the transformation laws for $\cD_A$ and $G_-$
\begin{subequations}
	\begin{align}
		\d_\s \cD_+ &= \hf \s \cD_+ + \cD_+\s M  ~, \\
		\d_\s \cD_{++} &= \s \cD_{++} - \ri \cD_+ \s \cD_+ + \cD_{++} \s M ~, \\
		\d_\s \cD_{--} &= \s \cD_{--} - \cD_{--} \s M ~, \\
		\d_\s G_- &= \frac 3 2 \s G_- + \frac 1 2 \cD_+ \cD_{--} \s~.
	\end{align}
\end{subequations}
These are exactly the super-Weyl transformations of the degauged geometry.

It should be noted that the above curved $ (1,0)$ superspace geometry was originally constructed in \cite{BMG,Gates:1986ez,EO}.


\section{Generalisations and future prospects}
\label{Section6}

Our approach to rigid $(p,q)$ superconformal symmetry was based on the use of real coordinates $z^{A} = (x^{a},\q^{+ \Io},\q^{- \Iu})$  to parametrise Minkowski superspace $\mathbb{M}^{(2|p,q)}$. The conformal Killing supervector fields were defined to satisfy
the equation \eqref{2.4}, which implied 
that the algebra of conformal Killing supervector fields of $\mathbb{M}^{(2|p,q)}$ is infinite dimensional. 
Its maximal finite-dimensional subalgebra is singled out by 
the conditions \eqref{Truncations}, and is isomorphic to  
 $\mathfrak{osp}(p|2;\mathbb{R}) \oplus \mathfrak{osp}(q|2;\mathbb{R})$.
The latter is the Lie algebra of the supergroup
${\sOSp}_0 (p|2; {\mathbb R} ) \times  {\sOSp}_0 (q|2; {\mathbb R} )$,
which is the superconformal group of the compactified Minkowski superspace \eqref{B.1}.

In the case that, say, $p$ is even, $p=2n$, the real Grassmann variables $\q^{+ \Io}$ can be replaced with complex ones, 
\bea
\q^{+ \Io} \to (\q^{+ i}~, \bar \q^+_i)~, \qquad \bar \q^+_i := \overline{\q^{+i}} ~, \quad 
i= 1, \dots , n~.
\eea   
At the same time, the real covariant derivatives $D_{+}^\OI$ should be replaced with complex ones
\bea
D^{\Io}_+ \to (D_{+ i}~, \bar D_+^i)~, \qquad \bar D_+^i := \overline{D_{+i}} ~,
\eea
which obey the algebra
\bea
\big \{ D_{+i} , \bar D_+^j \big \} = 2 \ri \d_i^j \partial_{++}~.
\eea

Then, equation \eqref{2.4} should be replaced with 
\begin{subequations}
\bea 
\label{complexSCKV}
[\xi , D_{+i} ] = - (D_{+i} \xi^{+ j}) D_{+ j} ~, \qquad
[\xi , D_-^{\Iu} ] = - (D_-^{\Iu} \xi^{- \Ju}) D_-^{\Ju} ~,
\eea
where $\xi$ takes the form
\bea
\xi = \xi^{a} \partial_{a} + \xi^{+ i} D_{+i} + \bar{\xi}^{+}_i \bar{D}_+^i + \xi^{- \UI} D_{-}^\UI = \bar{\xi}~.
\eea
\end{subequations}
One may then perform a similar analysis to that of section \ref{Section2} to obtain the corresponding superconformal algebra as a subalgebra of the $(p,q)$ super Virasoro algebra. We will not perform a complete analysis here and instead sketch the 
salient points.

The defining relations \eqref{complexSCKV} imply the master equations
\begin{align}
	\label{6.5}
	D_{+i} \xi^{--} = 0 ~, \qquad D_{-}^{\UI} \xi^{++}=0 ~, \qquad D_{+i} D_{+j} \xi^{++} = 0~,
\end{align}
and yield the following expressions for the spinor parameters
\begin{align}
	\xi^{+ i} = - \frac \ri 2 \bar D_{+}^i \xi^{++} ~, \qquad \xi^{- \UI} = - \frac \ri 2 D_{-}^\UI \xi^{--}~.
\end{align}
It should be noted that the final relation of \eqref{6.5} has the following non-trivial implications, depending on the value of $n$
\begin{subequations}
\begin{align}
	n = 2:& \quad \partial_{++} [D_{+i}, \bar{D}_{+}^i] \xi^{++} = 0 ~, \\
	n > 2:& \quad \partial_{++} [D_{+i}, \bar{D}_{+}^j] \xi^{++} = 0 ~.
\end{align}
\end{subequations}
In particular, it follows from the latter constraint that
\begin{align}
	\label{6.8}
	\partial_{++} \partial_{++} D_{+i} \xi^{++} = 0~,
\end{align}
and thus, for $n>2$, the vector $\xi^{++}$ encodes finitely many parameters. This is in contrast to the situation in section \ref{Section2}, where it was necessary to impose the conditions \eqref{Truncations}. This difference is a consequence of \eqref{complexSCKV} being more restrictive than \eqref{2.4}. For $n=1,2$ condition \eqref{6.8} must instead be imposed by hand.

Making use of \eqref{6.5}, the master equations \eqref{complexSCKV} may be written in the form
\begin{subequations}
	\begin{align}
		[\xi , D_{+i}] &= - \hf (\s[\xi] + K[\xi] + 2 \chi[\xi]) D_{+i} - \chi[\xi]_{i}{}^j D_{+j} ~, \\
		[\xi , D_-^{\Iu}] &= - \hf (\s[\xi] - K[\xi]) D_-^{\Iu} - \r[\xi]^{\Iu \Ju} D_{-}^{\Ju} ~,
	\end{align}
\end{subequations}
where $\s[\xi]$, $K[\xi]$ and $\r[\xi]^{\UI \UJ}$ were defined in \eqref{2.8} and we have made the definitions
\begin{align}
	\chi[\xi] := - \frac{\ri}{4n} [D_{+i}, \bar{D}_+^i] \xi^{++} ~, \qquad \chi[\xi]_i{}^j := - \frac{\ri}{4n} \Big ( [D_{+i}, \bar{D}_+^j] - \frac 1 n \d_i^j [D_{+k}, \bar{D}_+^k] \Big ) \xi^{++} ~. 
\end{align}
The former are constrained to satisfy \eqref{2.9}, while the new parameters obey
\begin{subequations}
	\begin{align}
		D_{+i} \chi[\xi] &= \frac{n-2}n D_{+i} \s[\xi] ~, \qquad  D_{-}^\UI \chi[\xi] = 0 ~, \\
		D_{+i} \chi[\xi]_j{}^k &= -2 \d_i^k D_{+j} \s[\xi] + \frac 2 n \d_j^k D_{+i} \s[\xi]~,
		\qquad D_{-}^\UI \chi[\xi]_j{}^k = 0~.
	\end{align}
\end{subequations}
One may then continue this analysis, keeping in mind the philosophy of section \ref{Section2}, to derive the superconformal algebra.

The resulting superconformal group for $n\neq2$ proves  to be 
 $\mathsf{SU} (1,1|n) \times  \mathsf{OSp}_0 (q|2; {\mathbb R})$, with 
$\sU(n)  \times \mathsf{SO}(q)$ being  its $R$-symmetry subgroup. 
The $n=2$ case is special since the diagonal $\sU(1)$ subgroup of 
$\mathsf{SU} (1,1|2 ) $
can be factored out, and the $R$-symmetry subgroup of the superconformal group becomes $\sSU(2) \times \mathsf{SO}(q)$. 
Now, the construction of conformal $(2n,q)$ superspace can be carried out by gauging 
the superconformal group $\mathsf{SU} (1,1|n ) \times  \mathsf{OSp}_0 (q|2; {\mathbb R} )$ for $n\neq 2$, or $\mathsf{PSU} (1,1|2 ) \times  \mathsf{OSp}_0 (q|2; {\mathbb R} )$ for $n=2$. The degauged version of this superspace may be denoted 
$\sU(n)  \times \mathsf{SO}(q)$ superspace, for $n\neq 2$, and $\sSU(2)  \times \mathsf{SO}(q)$ superspace, for $n=2$. Analogous considerations apply in the case that $q$ is even. In particular, for $p=q=4$, one can introduce conformal $(4,4)$ superspace as the gauge theory of $\mathsf{PSU} (1,1|2 ) \times \mathsf{PSU} (1,1|2 ) $. Its degauged version turns out to be the curved  $\sSU(2)  \times \sSU(2)$ superspace geometry
 introduced in \cite{TM}.\footnote{As pointed out in \cite{TM}, 
 the super-Weyl and local $R$-symmetry transformations can be used to partially fix the gauge freedom such that the resulting supergravity multiplet turns into the one proposed 
 many years ago in  \cite{Gates:1988ey, Gates:1988tn}.}
 It should be pointed out that the formulation for $\cN=(4,4)$ matter-coupled supergravity
 in $\sSU(2) \times \sSU(2)$ harmonic superspace was constructed in \cite{Bellucci:2000yx}. 
 We believe this formulation provides a solution to the torsion constraints 
for $\sSU(2)  \times \sSU(2)$ superspace \cite{TM} in terms of unconstrained prepotentials. 
It would be interesting to prove this conjecture.

 The $d=2$ superconformal groups were classified by 
 G\"unaydin, Sierra and Townsend \cite{GST} and have the structure 
 \bea
 \label{6.10}
  G = G_L \times G_R~,
 \eea
where $G_L$ and $G_R$ are simple supergroups. The supergroups  $G_L$ and $G_R$ can be any of the following: (i) $\sOSp (m|2;{\mathbb R})$; 
(ii) $\sSU(1,1|m)$, for $m\neq 2$, or $\mathsf{PSU}(1,1|2)$;
(iii) $\sOSp(4^*|2m )$; (iv) $\mathsf{G}(3)$; (v) $\mathsf{F}(4)$; and 
(vi) $\mathsf{D}^1(2,1,\a)$. 

Our paper has been devoted to conformal $(p,q)$ supergravity. Various versions of extended Poincar\'e supergravity may be obtained from $H_L \times H_R$ superspaces via coupling to compensating multiplets, following the universal approach advocated in \cite{KakuTownsend}.

Supersymmetric theories in AdS$_2$ have recently attracted much interest, see e.g.
\cite{Beccaria:2019dju} and references therein. Using our construction of $\mathsf{SO}(p) \times \mathsf{SO}(q)$ superspace, it is possible to work out the structure of AdS superspaces in two dimensions. This can be achieved by analogy with the derivation of $(p,q)$ superspace in three dimensions \cite{KLT-M12}. Specifically, two-dimensional AdS superspaces 
correspond to those supergravity backgrounds which 
satisfy the following requirements:

(i) the torsion and curvature tensors are Lorentz invariant;

(ii) the torsion and curvature tensors are covariantly constant.\\
Condition (i) means that 
\begin{align}
	X_{++}^{\OI \OJ} = 0~, \qquad X_{--}^{\UI \UJ} = 0~.
\end{align}
Condition (ii) is equivalent to the requirement
\begin{align}
\cD_A S^{\OI \UJ} = 0~,
\end{align}
which has nontrivial implications. This requirement and 
the relation \eqref{(p,q)algebra.b} give 
\bea
0= \{ \cD_{+}^{\OI}, \cD_{-}^{\UJ} \} S^{\OK \UL} = \d^{\OI \OK} S^{\OM \UJ} S^{\OM \UL} - \d^{\UL \UJ} S^{\OI \UM} S^{\OK \UM}~.
\eea 
The simplest solution for $p=q \equiv \cN$  is
\begin{align}
	S^{\OI \UJ} = S \d^{\OI \UJ}~, \qquad \cD_A S = 0~.
\end{align}
It corresponds to a special frame (or a special gauge condition) in which  
the left and right $R$-symmetry connections coincide 
\begin{align}
	\Phi_A^{\OI \OK} \d^{\OK \UJ} = \d^{\OI \UK} \Phi_A^{\UK \UJ} 
	~.
\end{align}
This means that the two types of $R$-symmetry indices turn into a single type, $\OI = \UI \equiv I$, and we stay with the diagonal subgroup of the $R$-symmetry group $\mathsf{SO}(\cN) \times \mathsf{SO}(\cN)$. The latter is generated by $J^{I J} = - J^{J I} = \mathfrak{L}^{IJ} + \mathfrak{R}^{IJ}$, which acts on isovectors as follows
\begin{align}
	J^{IJ} \chi^{K} = 2 \d^{K[I} \chi^{J]}~.
\end{align}
To summarise, the algebra of covariant derivatives for $\cN$-extended anti-de-Sitter superspace is
\begin{subequations} \label{AdSalgebra}
	\bea
	\{ \cD_{+}^{I}, \cD_{+}^{J} \} &=& 2 \ri \d^{I J} \cD_{++} ~, \quad \{ \cD_{-}^{I}, \cD_{-}^{J} \} = 2 \ri \d^{I J} \cD_{--} ~,  \\
	\{ \cD_{+}^{I}, \cD_{-}^{J} \} &=& - 4 \d^{I J} S M - 2 S J^{I J} ~, \\
	\big[ \cD_{+}^{I} , \cD_{--} \big]
	& = & - 2 \ri S \cD_{-}^{I}
	~, \quad 
	\big[ \cD_{-}^{I} , \cD_{++} \big]
	= 2 \ri S \cD_{+}^{I} ~,
	\\
	\big[ \cD_{++} , \cD_{--} \big]
	& = & 8 S^2 M
	~,
	\eea
\end{subequations}
where it should be understood that $J^{IJ}$ is not present for $\cN=1$. The AdS curvature $S$ is related to the scalar curvature by $\mathcal{R} = 16 S^2 < 0$.
The isometry supergroup of this $\cN$-extended AdS superspace is $\sOSp(\cN|2; {\mathbb R})$, see  \cite{GST} for the complete list of AdS$_2$ supergroups.  

The geometric structure of two-dimensional $(p,q)$ supersymmetric nonlinear $\s$-models is remarkably rich, see
\cite{Hull:1985jv, GHR, Hull, Lindstrom:2005zr, Hull:2018jkr} and references therein;
see also \cite{Lindstrom:2022lld} for a recent review. 
Rigid superconformal $\s$-models can be readily coupled to conformal supergravity. 
For non-superconformal $\s$-models, their uplift to curved superspace may be achieved by turning on a conformal compensator.

The formalism developed in this work may also be used to construct supersymmetric extensions of the Gauss-Bonnet invariant by a generalisation of four-dimensional logarithm construction of \cite{BdeWKL}. To this end, we consider a nowhere vanishing primary scalar (super)field $\varphi$ of non-zero dimension $\D$. From $\varphi$, one may construct the following primary descendants:
\begin{subequations} \label{6.200}
	\begin{align}
		\cN = (0,0):& \qquad \nabla_{++} \nabla_{--} \, \text{ln} \, \varphi ~, \label{6.200a}\\
		\cN = (1,0):& \qquad \nabla_{--} \nabla_{+} \, \text{ln} \, \varphi ~, \\
		\cN = (1,1):& \qquad \hf [ \nabla_{+} , \nabla_{-} ] \, \text{ln} \, \varphi ~.
		\label{6.200c}
	\end{align}
\end{subequations}
They can be used to define the (super)conformal functionals:
\begin{subequations}
\label{6.20}
\begin{align}
\cN = (0,0):& \qquad \cS_{(0,0)} = - \frac{1}{2\D} \int {\rm d}^2x \, e \, \nabla_{++} \nabla_{--} \, \text{ln} \, \varphi = - \frac{1}{8} \int {\rm d}^2x \, e \, \cR  ~, \label{6.20a}\\
\cN = (1,0):& \qquad \cS_{(1,0)} = - \frac{1}{2\D} \int {\rm d}^{(2|1,0)}z^{-} \, E \, \nabla_{--} \nabla_{+} \, \text{ln} \, \varphi = \int {\rm d}^{(2|1,0)}z^{-} \, E \, G_{-} ~, \\
\cN = (1,1):& \qquad \cS_{(1,1)} = - \frac{1}{4\D} \int {\rm d}^{(2|1,1)}z \, E \, [ \nabla_{+} , \nabla_{-} ] \, \text{ln} \, \varphi = \int {\rm d}^{(2|1,1)}z \, E \, S ~.
\label{6.20c}
\end{align}
\end{subequations}
Where in the latter expressions we have degauged the Lagrangian and then ignored all $\vf$-dependent surface terms. Remarkably, these functionals have proven to be independent of $\varphi$. Additionally, the first is simply the Gauss-Bonnet invariant, while the latter two are its simplest supersymmetric extensions. 

Using the primary (super)fields \eqref{6.200} allows us to introduce manifestly (super)conformal generalisations of the Fradkin-Tseytlin term in string theory \cite{Fradkin:1985ys}
\bea
S_{\rm FT}=  \frac{1}{4\p} \int {\rm d}^2x \, e \, \cR {\bm \F}~,
\eea
where $\bm \F$ denotes the dilaton field in the curved spacetime in which the string propagates. If we symbolically denote by $\O(\vf)$ any of the primaries in \eqref{6.200a} -- \eqref{6.200c} and by $\int \rd \m $ the integration measures in 
\eqref{6.20a} -- \eqref{6.20c}, then $I:= \int \rd \m \, \O(\vf)  {\bm \F}$ is invariant under the gauge group of conformal supergravity for any dimensionless primary scalar $\bm \F$. Here $\vf$ plays the role of a conformal compensator.  Choosing a (super-)Weyl gauge $\vf =1$ leads to standard expressions for the Fradkin-Tseytlin term and its supersymmetric extensions, modulo an overall numerical coefficient.

The analysis above may be extended to the $\cN=(2,2)$ case.
To this end, it is necessary to work in the complex basis of spinor covariant derivatives, which is obtained from \eqref{ComplexBasis} upon the replacement $\cD \rightarrow \nabla$. This allows us to define two types of constrained superfields, namely chiral superfields $\Phi$
\begin{subequations}
\label{constraints}
\begin{align}
	\bar{\nabla}_+ \Phi = 0 ~, \qquad \bar{\nabla}_- \Phi = 0~,
\end{align}
and twisted chiral superfields $\chi$
\begin{align}
	\bar{\nabla}_+ \chi = 0 ~, \qquad \nabla_- \chi = 0~,
\end{align}
\end{subequations}
where the Lorentz weights of $\Phi$ and $\chi$ are not indicated. Assuming that they are primary, their superconformal properties are related as follows:
\begin{subequations}
	\begin{align}
		q^L_\Phi &= \D_\Phi + \l_\Phi ~, \qquad q^R_\Phi = \D_\Phi - \l_\Phi~, \\
		q^L_\chi &= \D_\chi + \l_\chi  ~, \qquad q^R_\chi = \l_\chi - \D_\chi  ~.
	\end{align}
\end{subequations}
Here $q^L_\Phi$ is defined by $\ri \mathfrak{L}^{\overline{1} \overline{2}} \Phi = q^L_\Phi \Phi$ and similarly for $q^R_\Phi$. 

We now specify to primary Lorentz scalars $\Phi$ and $\chi$ of non-zero dimensions. From these superfields we may construct the primary descendants 
\bea
\cN = (2,2): \qquad 
\bar{\nabla}_+ \bar{\nabla}_- \, \text{ln} \, \bar{\Phi}~,
\qquad \bar{\nabla}_+ {\nabla}_- \, \text{ln} \, \bar{\chi}~, 
\eea
which are chiral and twisted chiral, respectively. They may be used to define the superconformal functionals
\begin{subequations}
	\label{6.23}
	\begin{align}
		\mathcal{S}^{\rm C}_{(2,2)} &= - \frac{1}{\D_\Phi} \int {\rm d}^{2}x \rd^2 \q \, \cE \, \bar{\nabla}_+ \bar{\nabla}_- \, \text{ln} \, \bar{\Phi} = \int {\rm d}^{2}x \rd^2 \q \, \cE \, \Xi^{\rm C} +\dots ~, \label{6.23a} \\
		\mathcal{S}^{\rm TC}_{(2,2)} &= \frac{1}{\D_\chi} \int {\rm d}^{2}x \rd {\q}^+ \rd \bar{\q}^- \, \mathfrak{E} \, \bar{\nabla}_+ {\nabla}_- \, \text{ln} \, \bar{\chi} = \int {\rm d}^{2}x \rd {\q}^+ \rd \bar{\q}^- \, \mathfrak{E} \, \Xi^{\rm TC}  +\dots ~, \label{6.23b}
	\end{align}
\end{subequations}
where $\cE$ and $\mathfrak{E}$ are appropriately defined measures for the chiral and twisted chiral subspaces, respectively, and we have made the definitions
\begin{subequations}
	\begin{align}
		\Xi^{\rm C} &:= S^{\overline{1} \underline{1}} + \ri S^{\overline{1} \underline{2}} + \ri S^{\overline{2} \underline{1}} - S^{\overline{2} \underline{2}}~, \quad \quad \quad \bar{\cD}_+ \Xi^{\rm C} = 0 ~, \quad \bar{\cD}_-\Xi^{\rm C} = 0~, \\
		\Xi^{\rm TC} &:= S^{\overline{1} \underline{1}} - \ri S^{\overline{1} \underline{2}} + \ri S^{\overline{2} \underline{1}} + S^{\overline{2} \underline{2}}~, \quad \quad \quad \bar{\cD}_+ \Xi^{\rm TC} = 0 ~, \quad {\cD}_-\Xi^{\rm TC} = 0~.
	\end{align}
\end{subequations}
It is not difficult to to show that the functionals \eqref{6.23a} and \eqref{6.23b} are independent of $\bar \F$ and $\bar \c$, respectively.  It may also be seen that these functionals are topological. Degauging the integrand in \eqref{6.23a} gives 
\bea
 \bar{\nabla}_+ \bar{\nabla}_- \, \text{ln} \, \bar{\Phi}  = - \D_\F \X^{\rm C} 
 +  \bar{\cD}_+ \bar{\cD}_- \, \text{ln} \, \bar{\Phi}  \equiv - \D_\F \X^{\rm C}  +\dots~,
\eea
where the ellipsis denotes a chiral superfield for which we do not yet have an explicit expression. Both functionals \eqref{6.23a} and \eqref{6.23b} define $\cN=(2,2)$ extensions of the Gauss-Bonnet invariant, eq. \eqref{6.20a}. Component analyses of these invariants will be given elsewhere.

Given primary dimensionless chiral $\J$ and twisted chiral $\S$  scalars, the following functionals 
\bea
\int {\rm d}^{2}x \rd^2 \q \, \cE \, \J \bar{\nabla}_+ \bar{\nabla}_- \, \text{ln} \, \bar{\Phi} ~,
\qquad
\int {\rm d}^{2}x \rd {\q}^+ \rd \bar{\q}^- \, \mathfrak{E} \, \S \bar{\nabla}_+ {\nabla}_- \, \text{ln} \, \bar{\chi} 
\eea
are superconformal invariants. They may be viewed as $\cN=2$ supersymmetric extensions  of the 
Fradkin-Tseytlin term.

The above $\cN=(2,2)$  constructions may also be generalised to the $\cN=(2,0)$ and $\cN=(2,1)$ cases. In both cases it is necessary to consider a scalar superfield $\F$ of dimension $\D$ which is chiral with respect to the left coordinates, $\bar{\nabla}_+ \Phi = 0$. Using $\Phi$, we construct the primary left chiral descendants
\begin{align}
	\cN = (2,0):& \qquad \bar{\nabla}_{+} \nabla_{--} \, \text{ln} \, \bar{\Phi} ~, \label{6.30} \\
	\cN = (2,1):& \qquad \bar{\nabla}_{+} \nabla_{-} \, \text{ln} \, \bar{\Phi} ~.
	\label{6.31}
\end{align}
They may be used to define the superconformal functionals
\begin{align}
	\cN = (2,0): \quad \mathcal{S}_{(2,0)} &= -\frac{1}{2\D} \int {\rm d}^{2}x \rd {\q}^+ \, \cE_L^{(2,0)} \, \bar{\nabla}_+ {\nabla}_{--} \, \text{ln} \, \bar{\Phi} \non \\
	\qquad &= 2\int {\rm d}^{2}x \rd {\q}^+ \, \cE_L^{(2,0)} \, G_{-} ~,  \label{6.29}\\
	\cN = (2,1): \quad \mathcal{S}_{(2,1)} &= \frac{1}{\sqrt{2}\D} \int {\rm d}^{2}x \rd {\q}^+ \rd {\q}^- \, \cE_L^{(2,1)} \, \bar{\nabla}_+ {\nabla}_- \, \text{ln} \, \bar{\Phi} \non \\
	\qquad &= \int {\rm d}^{2}x \rd {\q}^+ \rd {\q}^- \, \cE_L^{(2,1)} \, \Xi^{\rm LC} +\dots ~, \label{6.30}
\end{align}
where $\cE_L^{(2,0)}$ and $\cE_L^{(2,1)}$ are the appropriate integration measures, $G_-$ is defined in \eqref{5.33} and we have made the definition
\begin{align}
		\Xi^{\rm LC} &:= S^{\overline{1}} + \ri S^{\overline{2} } ~, \quad \quad \quad \bar{\cD}_+ \Xi^{\rm LC} = 0 ~.
\end{align}
Making use of the primary chiral descendants \eqref{6.30} and \eqref{6.31} allows us to define $(2,0)$ and $(2,1)$ supersymmetric analogues of the Fradkin-Tseytlin term. 
In both cases such invariants are associated with a primary dimensionless chiral scalar $\J$ and have the explicit form 
\begin{align}
	\cN = (2,0): \qquad & \int {\rm d}^{2}x \rd {\q}^+ \, \cE_L^{(2,0)} \,\J \bar{\nabla}_+ {\nabla}_{--} \, \text{ln} \, \bar{\Phi} \\
	\cN = (2,1): \qquad  & \int {\rm d}^{2}x \rd {\q}^+ \rd {\q}^- \, \cE_L^{(2,1)} \, \J\bar{\nabla}_+ {\nabla}_- \, \text{ln} \, \bar{\Phi} ~.
	\end{align}

Recently, new supertwistor formulations were discovered for 
conformal supergravity theories  in diverse dimensions  $3\leq d \leq 6$ \cite{HL20}.
It would be interesting to extend this approach to the $d=2$ case. 
\\

\noindent
{\bf Acknowledgements:}\\
We are grateful to Stefan Theisen for discussions and for pointing out an error in an earlier version of the manuscript. We thank Daniel Butter for a question that has led to 
the material in appendix \ref{AppendixC}.
The work of SK is supported in part by the Australian 
Research Council, project No. DP200101944.
The work of ER is supported by the Hackett Postgraduate Scholarship UWA,
under the Australian Government Research Training Program.


\appendix

\section{Conformal geometry in $d \geq 3$ dimensions} \label{AppendixA} 

This appendix is devoted to a brief review of 
conformal gravity in $d \geq 3$ dimensions as the gauge theory of the
conformal group $\sSO(d,2)$. This approach was pioneered in four dimensions in \cite{KTvN1}. 
Our discussion is aimed at elucidating the differences between the $d=2$ and $d>2$ cases. We closely follow the presentations given in  \cite{BKNT-M1,BKNT}.

The former is a gauge theory of the conformal
algebra $\mathfrak{so}(d,2)$, which is spanned by the translation ($P_a$), Lorentz ($M_{ab}$), dilatation ($\mathbb{D}$) and special conformal generators ($K^a$). Their non-vanishing commutation relations are
\begin{subequations}
	\label{A.1}
	\begin{align}
		[M_{ab} , M_{cd}] &= 2 \eta_{c[a} M_{b] d} - 2 \eta_{d [a} M_{b] c} ~, \\
		[M_{ab} , P_c ] &= 2 \eta_{c [a} P_{b]} ~, \quad [\mathbb D, P_a] = P_a ~, \\
		[M_{ab} , K_c] &= 2 \eta_{c[a} K_{b]}  ~, \quad  [\mathbb D, K_a] = - K_a ~, \\
		[K_a , P_b] &= 2 \eta_{ab} \mathbb D + 2 M_{ab} ~.
	\end{align}
\end{subequations}
It is convenient to group these generators into the two disjoint subsets $P_a$ and $X_{\underline a}$:
\begin{align}
	X_{\tilde a} = (P_a, X_{\underline a}) ~, \qquad X_{\underline a} = (M_{ab},\mathbb{D},K^a)~.
\end{align}
Then, the conformal algebra \eqref{A.1} may be rewritten as follows
\bsubeq
\begin{align}
	[X_{\underline{a}} , X_{\underline{b}} ] &= -f_{\underline{a} \underline{b}}{}^{\underline{c}} X_{\underline{c}} \ , \\
	[X_{\underline{a}} , P_{{b}} ] &= -f_{\underline{a} { {b}}}{}^{\underline{c}} X_{\underline{c}}
	- f_{\underline{a} { {b}}}{}^{ {c}} P_{ {c}} \label{A.2}
	\ .
\end{align}
\esubeq

\subsection{Gauging the conformal algebra in $d \geq 3$ dimensions}
\label{sectionA.1}

Let $\mathcal{M}^d$ be a $d$-dimensional spacetime, $d \geq 3$, parametrised by the local coodinates $x^m$, $m = 0,1, \dots, d-1.$ To gauge the conformal algebra \eqref{A.1} it is necessary to associate each non-translational generator $X_{\underline{a}}$ with a connection one-form $\o^{\underline{a}} = (\o,b,\mathfrak{f}_a)=\rd x^{m} \o_m{}^{\underline{a}}$ and with $P_a$ a vielbein one-form $e^a = \rd x^m e_m{}^a$, where it is assumed that $e:={\rm det}(e_m{}^a) \neq 0$, hence there exists a unique inverse vielbein $e_a{}^m$:
\begin{align}
	e_a{}^m e_m{}^b = \d_a{}^b~, \qquad e_m{}^a e_a{}^n=\d_m{}^n~.
\end{align}
The latter may be used to construct
the vector fields $e_a = e_a{}^m \pa_m $, which constitute a basis for the tangent space at each
point of $\mathcal{M}^{d}$. It may then be used to express the connection in the vielbein basis as
$\omega^{\underline{a}} =e^b\omega_b{}^{\underline{a}}$, 
where $\omega_b{}^{\underline{a}}=e_b{}^m\omega_m{}^{\underline{a}}$. 

The covariant derivatives
have the form\footnote{We adopt the convention where a factor of $1/2$ is inserted when performing a summations over pairs of antisymmetric indices.}
\be
\nabla_a =  e_a{}^m \partial_m - \hf \omega_a{}^{bc} M_{bc} - b_a \mathbb D - \mathfrak{f}_{ab} K^b ~.
\ee
We note that the translation generators $P_a$ do not appear in the covariant derivatives. Instead, we assume that they are replaced by $\nabla_a$ and obey the commutation relations:
\begin{align}
	[X_{\underline{a}} , \nabla_{{b}} ] &= -f_{\underline{a} { {b}}}{}^{\underline{c}} X_{\underline{c}}
	- f_{\underline{a} { {b}}}{}^{ {c}} \nabla_{ {c}} 
	\ .
\end{align}

By definition, the gauge group of conformal gravity is generated by local transformations of the form
\begin{subequations}\label{CGgentransmations}
	\bea
	\delta_{\mathscr K} \nabla_a &=& [\mathscr{K},\nabla_a] \ , \\
	\mathscr{K} &=& \xi^b \nabla_b +  \L^{\underline{b}} X_{\underline{b}}
	=  \xi^b \nabla_b + \hf K^{bc} M_{bc} + \s \mathbb{D} + \L_b K^b ~,
	\eea
\end{subequations}
where  the gauge parameters satisfy natural reality conditions. These gauge transformations act
on a conformal tensor field $\mathcal{U}$ (with its indices suppressed) as 
\bea 
\label{CGgenMattertfs}
\d_{\mathscr K} \mathcal{U} = {\mathscr K} \mathcal{U} ~.
\eea
Further, we will say that $\mathcal{U}$ is primary if (i) it is annihilated by the special
conformal generator, $K^a \mathcal{U} = 0$; and (ii) it is an eigenvector of $\mathbb D$. It will be said to have 
dimension $\D$ if $\mathbb D \cU = \D \cU$.

Amongst themselves, the covariant derivatives satisfy the following commutation relations
\be
[\nabla_a , \nabla_b] = -\cT_{ab}{}^c \nabla_c - \hf \cR(M)_{ab}{}^{cd} M_{cd}
- \cR(\mathbb{D})_{ab} \mathbb D - \cR(K)_{abc} K^c \ ,
\ee
where we have made the definitions:
\begin{subequations}
	\begin{align}
		\mathcal{T}_{ab}{}^c&=-\mathscr{C}_{ab}{}^{c}+2{\o}_{[ab]}{}^c+2{b}_{[a}\delta_{b]}{}^{c}~,\\
		\mathcal{R}(M)_{ab}{}^{cd}&=R_{ab}{}^{cd}+8\mathfrak{f}_{[a}{}^{[c}\delta_{b]}{}^{d]}~,\label{LorCur}\\ 
		\mathcal{R}(K)_{abc}&=-\mathscr{C}_{ab}{}^d\mathfrak{f}_{dc}-2{\o}_{[a|c|}{}^{d}\mathfrak{f}_{b]d}-2{b}_{[a}\mathfrak{f}_{b]c}+2e_{[a}\mathfrak{f}_{b]c}~, \label{2.19cc}\\
		\mathcal{R}(\mathbb{D})_{ab}&= -\mathscr{C}_{ab}{}^{c}{b}_c+4\mathfrak{f}_{[ab]}+2e_{[a}{b}_{b]}~,\\
		R_{ab}{}^{cd}&=-\mathscr{C}_{ab}{}^{f}{\o}_{f}{}^{cd}+2e_{[a}{\o}_{b]}{}^{cd}-2{\o}_{[a}{}^{cf}{\o}_{b]f}{}^{d}~. \label{RiemannB}
	\end{align}
\end{subequations}
Here $R_{ab}{}^{cd}$ is the curvature tensor\footnote{It should be emphasised that, owing to its dependence on the dilatation connection, our curvature tensor does not satisfy the Bianchi identity $R_{[abc]d}=0$ unless $b_a=0$, see the following subsection.} constructed from the Lorentz connection $\omega_a{}^{bc}$ and we have introduced the anholonomy coefficients $\mathscr{C}_{ab}{}^{c}$
\begin{align}
	[e_a,e_b] = \mathscr{C}_{ab}{}^{c} e_c ~.
\end{align}

In order for the above geometry to describe conformal gravity, it is necessary to impose certain covariant constraints such that the only independent geometric fields are the vielbein and dilatation connection. They are as follows:
\begin{subequations}
	\label{CC}
	\begin{align}
		\cT_{ab}{}^c &=0 ~, \label{CCa}\\
		\eta^{bd} \cR(M)_{abcd} &=0~. \label{CCb}
	\end{align}
\end{subequations}
The first constraint determines $\o_{a}{}^{bc}$ in terms of the vielbein and dilatation connection, while the second determines $\mathfrak{f}_{ab}$ to be
\be
\label{SCconn}
\mathfrak{f}_{ab} = - \hf P_{ab} = - \frac{1}{2 (d - 2)} R_{ab} + \frac{1}{4 (d - 1) (d - 2)} \eta_{ab} R ~,
\ee
where $P_{ab}$ is the Schouten tensor and we have defined
\be 
R_{ac} = \eta^{bd} R_{abcd} \ , \quad R = \eta^{ab} R_{ab} ~.
\ee
Inserting \eqref{SCconn} into \eqref{LorCur} leads to the result that $\cR(M)_{ab}{}^{cd}$ is exactly the Weyl tensor
\be
R(M)_{ab}{}^{cd} = C_{ab}{}^{cd} = R_{ab}{}^{cd} - 4 P_{[a}{}^{[c} \d_{b]}{}^{d]} ~,
\ee
which is a primary field of dimension 2
\begin{align}
	K_e C_{abcd} = 0 ~, \qquad \mathbb{D} C_{abcd} = 2 C_{abcd}~.
\end{align} 
It should be emphasised that, since $C_{abcd}$ is primary, it is independent of the dilatation connection $b_a$.

Next, it is necessary to analyse the Bianchi identity 
\be
[\nabla_a , [\nabla_b, \nabla_c]] + [\nabla_c , [\nabla_a, \nabla_b]] + [\nabla_b , [\nabla_c, \nabla_a]] = 0~,
\ee
which leads to the identities
\begin{subequations}
	\begin{align}
		\cR(\mathbb{D})_{ab} &= 0 ~, \\
		\cR(K)_{[abc]} &= 0 \ , \label{CGDBI1} \\ 
		C_{[abc]}{}^d &= 0 \ , \label{CGDBI2} \\ 
		\nabla_{[a} \cR(K)_{bc]}{}^d &= 0 \ , \label{CGDBI3} \\
		\nabla_{[a} C_{bc ]}{}^{de} - 4 \cR(K)_{[ab}{}^{[d} \d^{e]}_{c]} &= 0 \label{CGDBI4} \ . 
	\end{align}
\end{subequations}
In particular, constraint \eqref{CGDBI4} implies the important identity
\be 
\label{A.17}
\hf \nabla_c C_{ab}{}^{c e} + (d - 3) \cR(K)_{ab}{}^e - 2 \cR(K)_{c [a}{}^c \d^e_{b]} = 0 \ .
\ee
Thus, to continue our analysis it is necessary to consider the cases $d>3$ and $d=3$ separately.

For $d > 3$, it follows from \eqref{A.17} that the special conformal curvature takes the form
\begin{align}
	\cR(K)_{abc} = \frac{1}{2(d-3)} \nabla^d C_{abcd}~,
\end{align}
which implies that the algebra of covariant derivatives is
\begin{align}
	[\nabla_a , \nabla_b] = - \frac{1}{2} C_{abcd} M^{cd} - \frac{1}{2(d-3)} \nabla^d C_{abcd} K^c ~.
\end{align}
To conclude our discussion of the $d>3$ case, we list  the algebraic properties 
of $C_{abcd}$:
\bea
C_{abcd}= C_{[ab] [cd] } = C_{cd ab} ~, \qquad C_{a[bcd]} =0 ~,\qquad \eta^{bc} C_{abcd}=0~.
\eea

The $d = 3$ case is special because the Weyl tensor identically vanishes, $C_{abcd} = 0$. As a result, the algebra of covariant derivatives takes the form
\be
[\nabla_a , \nabla_b] = -\cR(K)_{abc} K^c ~,
\ee
and conformal geometry of spacetime is controlled by the primary field $\cR(K)_{abc}$. One can show that this field takes the form
\begin{align}
	\cR(K)_{abc} = - \cD_{[a} P_{b]c} = - \hf W_{abc} ~, \qquad \cD_a = e_a{}^m \partial_m - \o_{a}{}^{bc} M_{bc} ~,
\end{align}
where we have introduced the Lorentz covariant derivative\footnote{This definition of $\cD_a$ is valid for generic spacetime dimensions.} $\cD_a$ and the Cotton tensor $W_{abc}$. The latter proves to be a primary field of dimension $3$,
\begin{align}
	K_d W_{abc} = 0 ~, \qquad \mathbb{D} W_{abc} = 3 W_{abc}~.
\end{align}
It is useful to introduce its dual
\begin{align}
	W_{ab} = \hf \ve_{acd} W^{cd}{}_b ~,
\end{align}
which proves to be symmetric and traceless,
\begin{align}
	W_{ab} = W_{ba} ~, \qquad \eta^{ab} W_{ab} = 0~,
\end{align}
and also satisfies the conservation equation
\begin{align}
	\nabla^b W_{ab} = 0~.
\end{align}


\subsection{Degauging to Lorentzian geometry}

As mentioned above, the only independent geometric fields in this geometry (for $d \geq 3$) are the vielbein and dilatation gauge field. Actually, the latter proves to be a purely gauge degree of freedom. Specifically, it transforms algebraically under special conformal transformations \eqref{CGgentransmations}
\begin{align}
	\mathscr{K}(\L) = \L_a K^a \qquad \implies \qquad \d_{\mathscr{K}(\L)} b_a = - 2 \L_a~.
\end{align}
Hence, it is possible impose the gauge condition $b_a = 0$ at the cost of breaking special conformal symmetry.\footnote{This process is known as `degauging'.} As a result, the connection $\mathfrak{f}_{ab}$ is no longer required for the covariance of $\nabla_a$ under the residual gauge freedom and it may be manually extracted
\begin{align}
	\nabla_a = \cD_a - \mathfrak{f}_{ab} K^b = \cD_a + \hf P_{ab} K^b ~.
\end{align}
It may then be shown that the Lorentz covariant derivatives $\cD_a$ satisfy the algebra
\begin{align}
	[\cD_a , \cD_b] = - \hf \big( C_{ab}{}^{cd} + 4 P_{[a}{}^{c} \d_{b]}{}^{d} \big) M_{cd}~.
\end{align}

Next, it is important to describe the supergravity gauge freedom of this geometry, which corresponds to the residual gauge transformations of \eqref{CGgentransmations} in the gauge $b_a = 0$. These include local $\mathcal{K}$-transformations of the form
\begin{subequations}
	\label{A.30}
	\bea
	\delta_{\mathcal K} \cD_A &=& [\mathcal{K},\cD_A] \ , \\
	\mathcal{K} &=& \xi^b \cD_b + \hf K^{bc} M_{bc} ~,
	\eea
	which acts on tensor superfields $\mathcal{U}$ (with indices suppressed) as
	\begin{align}
		\d_\cK \cU = \cK \cU ~.
	\end{align}
\end{subequations}
The gauge transformations \eqref{A.30} prove to not be the most general conformal gravity gauge transformations preserving the gauge $b_a=0$.
Specifically, it may be shown that the following transformation also enjoys this property
\begin{align}
	\mathscr{K}(\s) = \s \mathbb{D} + \frac{1}{2} \nabla_b \s K^b \quad \Longrightarrow \quad \d_{\mathscr{K}(\s)} b_a = 0~,
\end{align}
where $\s$ is real but otherwise unconstrained. 

As a result, it is necessary to consider how this transformation manifests in the degauged geometry
\begin{align}
	\d_{\mathscr{K}(\s)} \nabla_a \equiv \d_{\s} \nabla_a = \d_\s \cD_a - \d_\s (\mathfrak{f}_{ab} K^b)~.
\end{align}
By a routine computation, we obtain
\begin{subequations}
	\begin{align}
		\d_\s \cD_a &= \s \cD_a + \cD^b \s M_{ba} ~, \\
		\d_\s C_{abcd} &= 2 \s C_{abcd} ~, \\
		\d_\s P_{ab} &= 2 \s P_{ab} - \cD_a \cD_b \s~,
	\end{align}
\end{subequations}
which are the standard Weyl transformations.


\section{Conformal $(1,0)$ superspace with non-vanishing curvature} \label{AppendixC}

In section \ref{Section4} we proposed conformal $(p,q)$ superspaces characterised by the relations \eqref{4.9}; all conformal curvatures were set to zero. This appendix is devoted to deriving  conformal $(1,0)$ superspace with non-vanishing curvature as an extension of the non-supersymmetric geometry \eqref{3.15}. 

Guided by the construction in higher dimensions \cite{ButterN=1,ButterN=2,BKNT-M1,BKNT-M3,BKNT}, we require that the covariant derivatives $\nabla_A = (\nabla_+ , \nabla_{++} , \nabla_{--})$ obey  constraints that are similar to the super Yang-Mills theory
\begin{align}
	\label{C.1}
	\big \{ \nabla_{+} , \nabla_{+} \big \} = 2 \ri \nabla_{++} ~, \qquad \big[ \nabla_{+} , \nabla_{--} \big] = \ri \mathscr{W}_{-} ~.
\end{align}
Here $\mathscr{W}_- = \mathscr{W}(X)_-{}^{\tilde A} X_{\tilde A}$, and $X_{\tilde A}$ denotes the generators of the superconformal algebra \eqref{4.1}. The Bianchi identities yield
\begin{align}
	\label{C.2}
	\big [ \nabla_{+} , \nabla_{++} \big ] = 0 ~, \qquad \big[ \nabla_{++} , \nabla_{--} \big] = \big \{ \nabla_{+} , \mathscr{W}_{-} \big \} ~.
\end{align}
Additionally, requiring consistency of \eqref{C.1} with the superconformal algebra leads to
\begin{align}
	\label{C.3}
	\big [K^{A}, \mathscr{W}_{-} \big \} = 0 ~, \qquad \big[\mathbb{D} , \mathscr{W}_- \big] = \frac 3 2 \mathscr{W}_- ~, \qquad \big[M , \mathscr{W}_- \big] = - \frac 1 2 \mathscr{W}_- ~.
\end{align}

Hence, we constrain $\mathscr{W}_-$ to be
\begin{align}
	\label{C.4}
	\mathscr{W}_{-} = \chi_{+} K^{++} + \psi_{---} K^{--}~,
\end{align} 
where $\chi_{+}$ and $\psi_{---}$ are primary superfields of dimension $5/2$. At the component level, they contain the conformal curvatures $W_{++}$ and $W_{--}$ \eqref{3.15}
\begin{align}
	\label{C.5}
	W_{++} = \nabla_{+} \chi_{+} |_{\q^+ = 0} ~, \qquad W_{--} = \nabla_{+} \psi_{---} |_{\q^+ = 0}~.
\end{align}
Finally, inserting \eqref{C.4} into \eqref{C.2}, we obtain the commutator of vector derivatives
\begin{align}
	\big[ \nabla_{++} , \nabla_{--} \big] = \ri \chi_+ S^+ + (\nabla_+ \chi_+) K^{++} + (\nabla_+ \psi_{---}) K^{--}~.
\end{align} 

We note that, according to \eqref{SUGRAtransmations}, the dilatation connection transforms algebraically under infinitesimal special superconformal gauge transformations \eqref{4.1}. This allows us to fix the gauge $B_A=0$ at the expense of this symmetry. In this gauge the special conformal connection $\mathfrak{F}_{AB}$ is 
not required for the covariance of $\nabla_A$ 
and may
be separated
\begin{align}
	\nabla_A = \cD_A - \mathfrak{F}_{AB} K^B~.
\end{align}
Here the degauged covariant derivative $\cD_A$ involves only the Lorentz connection and obeys the algebra \eqref{(1,0)algebra}. The connection $\mathfrak{F}_{AB}$ was described in \eqref{5.34a} and \eqref{5.34b}. 
The constraints \eqref{5.34c} are now replaced with 
\begin{align}
	\psi_{---} = \cD_{--} G_{-} - \cD_+ \mathfrak{F}_{--,--} ~, \qquad \chi_+ = - \cD_{++} G_- + \cD_{--} \mathfrak{F}_{+,++}~.
\end{align}
These relations determine the curvature tensors $\j_{---}$ and $\c_+$ in terms of $\cD_A$ and $\mathfrak{F}_{AB}$.
Keeping \eqref{C.5} in mind, it is clear that this is a $(1,0)$ extension of \eqref{323}. In particular, for vanishing $\psi_{---}$  and $\chi_+$, the connections $\mathfrak{F}_{--,--}$  and $\mathfrak{F}_{+,++}$ are  non-local functions of the supergravity multiplet.


\section{Compactified Minkowski superspace}\label{AppendixB}

Superconformal groups \eqref{6.10} do not act on Minkowski superspace,
since the special conformal and $S$-supersymmetry transformations are singular at some points. However, there exists a well defined action of \eqref{6.10} on a compactified $(p,q)$ Minkowski superspace $\overline{\mathbb M}^{(2|p,q)}$, for some $p,q$. In general, $\overline{\mathbb M}^{(2|p,q)}$ has the form
\bea
\overline{\mathbb M}^{(2|p,q)} = S^{1|p}_L \times S^{1|q}_R~,
\label{B.1}
\eea
where the bosonic body of $S^{1|n}$ is a circle $S^1$. The left superspace $S^{1|p}_L$ is a homogeneous space of the subgroup $G_L$
of \eqref{6.10}, and similarly in the right sector.

In a recent paper \cite{KT-M2021}, $\overline{\mathbb M}^{(2|p,q)}$ was realised
as a homogeneous space of the superconformal group
${\sOSp}_0 (p|2; {\mathbb R} ) \times  {\sOSp}_0 (q|2; {\mathbb R} )$.
Here we present a different construction for the case that $p$ is even, $p =2n$.
Specifically, we describe
$\overline{\mathbb M}^{(2|2n,q)}$
as a homogeneous space of the superconformal group
\bea
G=G_L\times G_R = \mathsf{SU} (1,1|n )
\times  {\sOSp}_0 (q|2; {\mathbb R} )~.
\eea

We begin by describing the action of $\mathsf{SU} (1,1|n )$ on $S^{1|2n}$.
The supergroup $\mathsf{SU} (1,1|n )$ is spanned by supermatrices of the form
\bea
g \in \sSL(2 | n;{\mathbb C} ) ~, \qquad
g^\dagger \,\O \,g = \O~,  \qquad
\O=
\left(
\begin{array}{cc|c}
	1&  0 ~&0\\
	0 &-1 ~&0\\
	\hline
	0 & 0& {\mathbbm 1}_{n}
\end{array}
\right) ~.
\eea
This supergroup naturally acts on the space of even supertwistors ${\mathbb C}^{2|n}$
\bea
X = \left(
\begin{array}{c}
	z \\
	w\\
	\hline
	\vf^i
\end{array}
\right) ~,\qquad i=1, \dots , n~,
\eea
where $z,w$ are complex bosonic variables, and $\vf^i$ complex Grassmann variables.
We identify $S^{1|2n}$ with the space of null lines in ${\mathbb C}^{2|n}$.
By definition,
a null supertwistor $X$ is characterised by the conditions
\bea
X^\dagger \O X=0~, \qquad  \left(
\begin{array}{c}
	z \\
	w
\end{array}
\right) \neq 0~.
\eea
Two null supertwistors $X$ and $X'$ are said to be equivalent if
\bea
X' = {\mathfrak c} X~, \qquad {\mathfrak c} \in {\mathbb C}\setminus \{0\}~.
\label{B.6}
\eea
Any equivalence class in the set of null supertwistors is called a null line. Given a null supertwistor $X$ both bosonic components $z$ and $w$ are non-zero.
Making use of the equivalence relation \eqref{B.6} allows us to choose, for each null line,
a representative
\bea
X = \left(
\begin{array}{c}
	z \\
	1\\
	\hline
	\vf^i
\end{array}
\right) ~,\qquad |z|^2 = 1 -\vf^\dagger \vf~,
\eea
which
is uniquely defined for the null line under consideration. It is seen that the quotient space is $S^{1|2n}$.

In order to make contact to ordinary Minkowski superspace, it is useful to switch to a different parametrisation of $\sSU(1,1|n)$ and the associated supertwistor space.
Let us introduce the supermatrix
\bea
\S= \frac{1 }{ \sqrt{2} }
\left(
\begin{array}{cr|c}
	{1}  ~ & - {1} ~& 0\\
	1 ~&    1 ~& 0  \\
	\hline
	0 & 0 ~ & \sqrt{2} \,{\mathbbm 1}_{n}
\end{array}
\right)~, \qquad \S^\dagger \S= {\mathbbm 1}_{2+n}~,
\eea
and associate with it the following similarity transformation:
\bea
g ~& \to & ~ \hat{ g} = \S \, g\, \S^{-1} ~, \quad g \in \sSU(1,1|n)~;
\qquad
X ~ \to  ~ \hat{X} = \S \, T~, \quad X \in {\mathbb C}^{2|n}~.
\label{sim2}
\eea
The supertwistor metric $\O$ turns into
\bea
\hat{ \O} =  \S \, \O\, \S^{-1}
=
\left(
\begin{array}{cc|c}
	0&  {1} ~&0\\
	{1} &0 ~&0\\
	\hline
	0 & 0& {\mathbbm 1}_{n}
\end{array}
\right) ~.
\eea
In the new frame, it is not guaranteed that both bosonic components $\hat z$ and $\hat w$ of  a null supertwistor $\hat X$ are non-zero. However, at least one of $\hat z$ and $\hat w$ is non-vanishing, and we can introduce an open subset of $S^{1|p}$ which is parametrised by null supertwistors of the form
\bea
\hat{X} = \left(
\begin{array}{c}
	1 \\
	- \ri {\bm x}^{++} \\
	\hline
	\sqrt{2}\q^{+i}
\end{array}
\right) ~, \qquad {\bm x}^{++} - \bar{\bm x}^{++} = 2\ri \bar \q^+_i \q^{+i} ~,
\quad \bar \q^+_i := \overline{\q^{+ i}}~.
\eea
The constraint on ${\bm x}^{++}$ is solved by
\bea
{\bm x}^{++} = {x}^{++} + \ri \bar \q^+_i \q^{+i} ~, \qquad
\overline{ x^{++}} = x^{++}~.
\eea
The variables ${\bm x}^{++}$ and $\q^{+i} $ parametrise a chiral subspace of $\mathbb{M}^{(2|2n,q)}$. To deduce the superconformal transformations of this subspace it is necessary to act on $\hat{X}$ with a generic group element $\hat{g} \in \sSU(1,1|n)$:
\begin{align}
	\hat{g} = \re^{\L_L} ~, \qquad \L_L = 
	\left(
	\begin{array}{cc|c}
		-\hf(\s + K) - \frac{\ri n }{n-2} \chi&  {\ri b_{++}} ~&\sqrt{2} \eta_{+j}\\
		{- \ri a^{++}} & \hf(\s + K) - \frac{\ri n }{n-2} \chi ~&\sqrt{2} \bar{\epsilon}^+_{j}\\
		\hline
		\sqrt{2} \epsilon^{+i} & \sqrt{2} \bar{\eta}^{i}_+& \l^{i}{}_j- \frac{2 \ri n }{n-2} \chi \d^{i}_j
	\end{array}
	\right)~,
\end{align}
where all scalar and vector parameters are real and 
\begin{align}
	\l^{\dagger} = - \l ~, \qquad {\rm tr} \ \l = 0~.
\end{align}
Taking the parameters to be small, one may show that the most general infinitesimal superconformal transformations on this subspace are:
\begin{subequations}
	\label{B.13}
	\begin{align}
		\d \bm{x}^{++} &= (\s + K) \bm{x}^{++} + a^{++} + 2 \ri \bar{\epsilon}_i^+ \q^{+i} - \bm{x}^{++} b_{++} \bm{x}^{++} - 2\bm{x}^{++} \eta_{+i} \q^{+i} ~, \\
		\d \q^{+i} &= \hf(\s+K) \q^{+i} -\frac{\ri n \chi}{n-2} \q^{+i} + \epsilon^{+i} + \l^{i}{}_j \q^{+j} - \q^{+i} b_{++} \bm{x}^{++} \non \\
		&\phantom{=}- \ri \bar{\eta}_{+}^i \bm{x}^{++}
		-2 \q^{+i} \eta_{+j} \q^{+j}  ~.
	\end{align}
\end{subequations}
The constant bosonic parameters in \eqref{B.13} correspond to dilatations $(\s)$, Lorentz transformations $(K)$, spacetime translations $(a^{++})$, special conformal transformations $(b_{++})$, chiral transformations $(\chi)$ and $\sSU(n)$ rotations $(\l^{i}{}_{j})$. The constant fermionic parameters correspond to $Q$-supersymmetry $(\epsilon^{+i})$ and $S$-supersymmetry $(\eta_{+ j})$ transformations.

Next, we consider the action of $\mathsf{OSp}_0 (q|2;\mathbb{R})$ on $S^{1|q}$, see \cite{KT-M2021} for more details.
This supergroup is spanned by supermatrices of the form
\bea
h \in \sSL(2 | q;{\mathbb R} ) ~, \qquad
h^{\rm sT} \, \mathbb{J} \,h = \mathbb{J} ~,  \qquad
\mathbb{J}=
\left(
\begin{array}{cc|c}
	0&  1 ~&0\\
	-1 & 0 ~&0\\
	\hline
	0 & 0& {\ri \mathbbm 1}_{n}
\end{array}
\right) ~,
\eea
and naturally acts on the space of even supertwistors ${\mathbb R}^{2|q}$
\bea
Y = \left(
\begin{array}{c}
	a \\
	b\\
	\hline
	\s^{\UI}
\end{array}
\right) ~,\qquad \UI=\underline{1}, \dots , \underline{q}~,
\eea
where $a,b$ denote real bosonic variables which are not both zero, and are $\s^\UI$ real Grassmann variables. Two supertwistors $Y$ and $Y'$ are equivalent if $Y' = \g Y$, where $\g \in \mathbb{R} \setminus \{ 0 \}$. Assuming that $a \neq 0$, we can choose the representative
\bea
Y = \left(
\begin{array}{c}
	1 \\
	-x^{--}\\
	\hline
	\ri \q^{-\UI}
\end{array}
\right) ~,
\eea
where $(x^{--},\q^{-\UI})$ constitute inhomogeneous coordinates for $S^{1|q}$. To deduce the superconformal transformations of this subspace it is necessary to act on $Y$ with a generic group element $h \in {\sOSp}_0 (q|2; {\mathbb R} )$
\begin{align}
	h = \re^{\L_R} ~, \qquad \L_R = 
	\left(
	\begin{array}{cc|c}
		-\hf(\s - K) &  {- b_{--}} ~& - \eta_-^{\UJ}\\
		{- a^{--}} & \hf(\s - K) &\sqrt{2} {\epsilon}^{- \UJ}\\
		\hline
		\ri \epsilon^{- \UI} & \ri {\eta}^{\UI}_-& \r^{\UI \UJ}
	\end{array}
	\right)~.
\end{align}
Here all parameters are real and $\r^{\UI \UJ} = - \r^{\UJ \UI}$. Taking the parameters to be small, it may be shown that the most general infinitesimal superconformal transformations on this subspace are:
\begin{subequations}
	\label{B.17}
	\begin{align}
		\d x^{--} &= (\s - K) x^{--} + a^{--} - x^{--} b_{--} x^{--} + \ri \epsilon^{- \UI} \q^{- \UI} + \ri x^{--} \eta_-^{\UI} \q^{- \UI}  ~, \\
		\d \q^{- \UI} &= \hf (\s - K) \q^{- \UI} - \q^{- \UI} b_{--} x^{--} + \epsilon^{- \UI} + \r^{\UI \UJ} \q^{- \UJ} - \eta_-^{\UI} x^{--} - \ri \q^{- \UI} \eta_-^{ \UJ} \q^{- \UJ}~.
	\end{align}
\end{subequations}
The constant bosonic parameters in \eqref{B.17} correspond to dilatations $(\s)$, Lorentz transformations $(K)$, spacetime translations $(a^{--})$, special conformal transformations $(b_{--})$ and $\sSO(q)$ rotations $(\r^{\UI \UJ})$. The constant fermionic parameters correspond to $Q$-supersymmetry $(\epsilon^{-\UI})$ and $S$-supersymmetry $(\eta_-^{\UI})$ transformations. We emphasise that the parameters $\s$ and $K$ in \eqref{B.13} and \eqref{B.17} are the same.


\begin{footnotesize}

\end{footnotesize}


\end{document}